\documentclass[12pt]{article}
\usepackage{amsmath, amsthm, amssymb}
\usepackage{setspace}
\usepackage{multirow}
\usepackage{graphicx}
\usepackage{apacite}
\usepackage{natbib}
\usepackage{color}
\usepackage{subcaption}
\usepackage{hyperref}
\usepackage{xr}

\newcommand{\uA}       {\mbox{\boldmath$A$}}

\newcommand{\uD}       {\mbox{\boldmath$D$}}

\newcommand{\uE}       {\mbox{\boldmath$E$}}

\newcommand{\uI}       {\mbox{\boldmath$I$}}

\newcommand{\uR}       {\mbox{\boldmath$R$}}
\newcommand{\ur}       {\mbox{\boldmath$r$}}
\newcommand{\uS}       {\mbox{\boldmath$S$}}

\newcommand{\uw}       {\mbox{\boldmath$w$}}
\newcommand{\uX}       {\mbox{\boldmath$X$}}
\newcommand{\ux}       {\mbox{\boldmath$x$}}
\newcommand{\uY}       {\mbox{\boldmath$Y$}}
\newcommand{\uy}       {\mbox{\boldmath$y$}}
\newcommand{\uZ}       {\mbox{\boldmath$Z$}}
\newcommand{\uz}       {\mbox{\boldmath$z$}}

\newcommand{\ualpha}       {\mbox{\boldmath$\alpha$}}
\newcommand{\ubeta}       {\mbox{\boldmath$\beta$}}
\newcommand{\ugamma}       {\mbox{\boldmath$\gamma$}}

\newcommand{\udelta}       {\mbox{\boldmath$\delta$}}

\newcommand{\utheta}       {\mbox{\boldmath$\theta$}}

\newcommand{\ulambda}       {\mbox{\boldmath$\lambda$}}
\newcommand{\umu}       {\mbox{\boldmath$\mu$}}

\newcommand{\uSigma}       {\mbox{\boldmath$\Sigma$}}
\newcommand{\usigma}       {\mbox{\boldmath$\sigma$}}

\newcommand{\uPsi}       {\mbox{\boldmath$\Psi$}}

\newcommand{\uzero}       {\mbox{\boldmath$0$}}
\newcommand{\uone}       {\mbox{\boldmath$1$}}

\newcommand{\utX} {\tilde{\boldsymbol{\hspace{-2pt}X}}}

\newcommand{\utS} {\tilde{\boldsymbol{\hspace{-2pt}S}}}

\newcommand{\utY} {\tilde{\boldsymbol{\hspace{-2pt}Y}}}

\newcommand{\ucX} {\check{\boldsymbol{\hspace{-0pt}X}}}
\newcommand{\ucy} {\check{\boldsymbol{\hspace{-0pt}y}}}

\title{Bayesian semiparametric analysis of multivariate continuous responses, with variable selection}

\author{Georgios Papageorgiou and Benjamin C. Marshall\\
	Department of Economics, Mathematics and Statistics\\
	Birkbeck, University of London, UK\\
	g.papageorgiou@bbk.ac.uk
}

\begin{document}
	\maketitle
	
	\begin{center}
		\emph{Abstract}
	\end{center}

This article presents an approach to Bayesian semiparametric inference for Gaussian multivariate response regression. We are motivated by various small and medium 
dimensional problems from the physical and social sciences. The statistical challenges revolve around dealing with the unknown mean and variance functions 
and in particular, the correlation matrix. To tackle these problems, we have developed priors over the smooth functions and a Markov chain Monte Carlo 
algorithm for inference and model selection. Specifically, Dirichlet process mixtures of Gaussian distributions are used as the basis for a cluster-inducing 
prior over the elements of the correlation matrix. The smooth, multidimensional means and variances are represented using radial basis function expansions. The 
complexity of the model, in terms of variable selection and smoothness, is then controlled by spike-slab priors. A simulation 
study is presented, demonstrating performance as the response dimension increases. Finally, the model is fit to a number of real world datasets.
An R package, scripts for replicating synthetic and real data examples, and a detailed description of the MCMC sampler are available in the supplementary 
materials online.\\
\\
\emph{Keywords}: Clustering; Covariance matrix models; Model averaging; Multivariate response regression; Seemingly unrelated regression models; Semiparametric regression
                  
\section{Introduction}

Many systems are too complex to be adequately described by a single response variable. 
For example, in medical investigations, understanding how the body reacts to certain drugs requires multiple blood tests. 
Similarly, in the social sciences, multiple exams are needed in order to build a complete picture of a student's academic ability.
Scientific investigations into these systems therefore produce multiple outcome variables. 
Typically the outcomes are correlated and ignoring this correlation can result in loss of optimality. 
Multivariate response models are needed for the analysis of data arising from these and many other experimental setups. 
Our main goal here is to develop Bayesian multivariate response models for continuous responses with nonparametric models for the mean vectors and covariance matrices, 
assuming a multivariate Gaussian likelihood.

Modelling unconstrained means nonparametrically, as general functions of the covariates, is straight forward and by now fairly standard. 
In the work that we present here, nonparametric effects are represented as linear combinations of radial basis functions. Generally, our approach is to utilize a large number of basis 
functions because this enables flexible estimation of true effects that are locally adaptive. Potential over-fitting is mitigated by utilizing spike-slab priors for variable selection and regularization 
(see e.g.  \citet{ohara2009} for a review on variable selection methods).

Modelling covariance matrices nonparametrically is not as straight forward as modelling the means, due the positive definiteness constraint that complicates matters. To overcome this constraint and model the elements of the covariance matrix in terms of regressors, a first, necessary step is to decompose the covariance matrix $\uSigma$ into a product of matrices. Such decompositions include the spectral and Cholesky, and variations of the latter. \citet{PinheiroB96} review the spectral and Cholesky decompositions with several different parametrisations. Based on the spectral decomposition and the matrix logarithmic transformation, \citet{Chiu96} model the structure of a covariance matrix in terms of explanatory variables. \citet{Poura1999} and \citet{Chen03} describe two modifications of the Cholesky decomposition that result in statistically meaningful, unconstrained reparametrisation of the covariance matrix, provided that there is a natural ordering in the responses \citep{Poura07}, as it happens in longitudinal studies, where the time of observation provides this natural ordering.   

The spectral and the modified Cholesky decompositions, outside the context of longitudinal studies, lack simple statistical interpretation, making it difficult for practitioners to incorporate  prior beliefs into the model. A decomposition, however, that is statistically simple and intuitive, comes from the separation strategy of \citet{Barnard00}, according to which $\uSigma$ is separated into a diagonal matrix of variances $\uS$ and a correlation matrix $\uR$, $\uSigma = \uS^{1/2} \uR \uS^{1/2}$. This decomposition makes it easy to model the variances in terms of covariates as the only constrained on them is the positiveness. Here we use a $\log$-link and linear predictors that are constructed in the same way as for the mean parameters. 

\citet{Chan06} describe several reasons why allowing the variances to be general functions of the covariates is meaningful. First, prediction intervals obtained from heteroscedastic regression models can be more realistic than those obtained by assuming constant error variance, or as \citet{muller2013} put it, it can result in more honest representation of uncertainties. Second, it allows the practitioner to examine and understand which covariates drive the variances, and in the multivariate response case, examine if the same or different subsets of covariates are associated with the variances of the responses. Third, modelling the variances in terms of covariates results in more efficient estimation of the mean functions. 
Lastly, it produces more accurate standard errors for the estimates of unknown parameters. 

Our approach for variable selection and model averaging can be thought of as a generalization of the approach of \citet{george93} who describe methods for univariate linear regression and the approach of \citet{Chan06} and \citet{pap18} who focused on methods for flexible mean and variance modelling for a 
single response. The current article is a generalization of the work of \citet{Chan06} and \citet{pap18} from univariate to multivariate responses. Whereas in the univariate 
case one has to fit a single smooth mean and a single smooth variance function, in the multivariate case, multiple  such functions have to be fit. However, the representation of these functions, and their prior distributions, are constructed in the same way as in the univariate case. The most important challenge that one has to face when dealing with multivariate regression is modelling the correlation matrix and sampling from its posterior. In this article, we discuss three intuitive correlation matrix priors and strategies for posterior sampling.
In addition, we develop an efficient stochastic search variable selection algorithm by using Zellner's g-prior \citep{AZ} that allows integrating out the regression coefficients in the mean function. Further, in our Markov chain Monte Carlo (MCMC) algorithm, 
we generate the variable selection indicators in blocks \citep{Chan06, pap18} and choose the MCMC tuning parameters adaptively \citep{Roberts2001c}. 

Of course, the separation of the variances from the correlations alone does not solve the problem of positive definiteness, as the constraint has now been transferred from the covariance matrix $\uSigma$ to the correlation matrix $\uR=\{r_{kl}\}$. Here, we place a normal prior on the Fisher's $z$ transformation of the nonredundant elements of $\uR$, $\log\{(1+r_{kl})/(1-r_{kl})\}/2 \sim N(\mu_R,\sigma^2_R) I[\uR \in \mathcal{C}]$, where $\mathcal{C}$ denotes the space of correlation matrices and $I[.]$ denotes the indicator function that restricts the range of the correlations and induces dependence among them \citep{DanielsKass99}. We rely on the `shadow prior' of \citet{Liechty} to maintain positive definiteness. The model is intuitive and easy to interpret, allowing practitioners to represent their substantive prior knowledge.  

However, the normal model for the correlations is quite restrictive, and this can have a negative impact on the estimated correlations, especially in small samples \citep{DanielsKass99}. Here, to achieve a nonparametric model for the correlation matrix, we consider mixtures of normal distributions $\log\{(1+r_{kl})/(1-r_{kl})\}/2 \sim \sum_{h} \pi_h N(\mu_{R,h},\sigma^2_R) I[\uR \in \mathcal{C}]$
for the transformed $r_{kl}$. This is in the spirit of the `grouped correlations model' of \citet{Liechty} who also propose 
a `grouped variables model'. The latter clusters the variables instead of the correlations and it is more structured than the nonparametric grouped correlations model. Here, we consider both the grouped correlations and variables models.    
   
In what follows, we work with generic Dirichlet process \citep{Fer73} mixtures of normal distributions for the correlations, utilizing the stick breaking construction \citep{sethuraman}. However, one of the attractive features of the grouped correlations and variables models is that they allow the researcher to represent prior information and beliefs about the strength of correlations among variables and the general structure of the correlation matrix. See \citet{Liechty} and \citet{TsayPoura17} for examples on structured correlation matrices.

Our work is related to two further strands of the literature. The first one is known as `seemingly unrelated regressions' (SUR) and it originates from the work of \citet{Zellner62}. The second one is known as `generalized additive models for location, scale and shape' (GAMLSS) and it originates from the work of \citet{RS05}.

Concerning SUR, \citet{Zellner62} showed how efficiency gains can be achieved by simultaneous estimation of linear regression equations, accommodating potentially correlated error terms. This gain in efficiency, measured in terms of reduction in the variance of the estimates of regression coefficients, can be substantial when the correlations among the error terms are high and covariates in different regression equations are not highly correlated. As the methodology presented in this article is a Bayesian semi-parametric version of Zellner's model, similar gains are to be expected from our approach too, and these are investigated in a simulation study presented in Section \ref{sim}.

GAMLSS, and the Bayesian analogue termed as BAMLSS \citep{Umlauf17}, provides a general framework for the analysis of data in a very wide class of univariate distributions, utilizing flexible models for the parameters of the response distribution. The popularity of these methods stems from the fact that for most realistic problems, the assumption that the parameters are linearly dependent on the covariates, or even constant (as in homoscedastic regression), is not tenable. Applying this level of regression flexibility to multivariate response models is currently an active area of research. \citet{SMITH2000257} implemented the multivariate normal regression model with smooth additive terms in the mean function and with homoscedastic errors.  \cite{KleinKneib} present applications of the GAMLSS framework to bivariate regression with normal and $t$-distributed errors, and on Dirichlet regression. \cite{Klein2016} used copulas in bivariate response models, relating the parameters of the marginals and those of the dependence structure to additive predictors. Here, we focus attention to models with Gaussian errors and we develop a fully multivariate model with nonparametric models for the means, the variances and the correlation matrix, with automatic variable selection based on spike-slab priors. 

The remainder of this article is arranged as follows. Section \ref{modes} develops the proposed model in more detail. Posterior sampling is discussed in Section \ref{postsamp}. Section \ref{sim} presents results from a simulation study that examines the efficiency gains one may have when fitting multivariate models instead of univariate ones. We also look into the performance of the method in automatically choosing the appropriate level of function complexity.
Lastly, the simulation study reports the run-times needed to fit the described models. 
In Section \ref{applications} we present two applications of the model. In the first application, the objective is to understand how the human body responds to a drug overdose. This is a common setting, where there are multiple outcomes each depending on a number of covariates. 
The statistical task is to understand how responses and covariates relate to each other. The second application is taken from the social sciences. 
Statistically, the problem can be seen as graphical modelling, where the conditional independence properties of the inverse covariance matrix are combined with flexible regression modelling. The article concludes with a brief discussion. All the methods we used in this article are freely available in the R package BNSP \citep{bnsp}. 

\section{Multivariate response model}\label{modes}

Let $\uy_{i} = (y_{i1},\dots,y_{ip})^{\top}$ denote a $p$-dimensional response vector and $\ux_i$ and $\uz_i$ 
denote covariate vectors, observed on the $i$th sampling unit, $i=1,\dots,n$.
Below we detail how the mean and covariance matrix of $\uy_i$ are modelled in terms of  
covariates. Here, $\ux_i$ denotes the vector of covariates for the mean and $\uz_i$ that for the covariance. 
Hence, we do not assume the covariates for the mean are necessarily the same as those for the covariance.
Typically, $\uz_i$ is a subset of $\ux_i$.

The model for the mean of the $j$th response, $j=1,\dots,p$, is expressed as 
\begin{equation}
E(Y_{ij}) = \mu_{ij} = \mu(\ux_i,\ubeta_j^{\ast}) = \beta_{0j} + \ux_i^{\top} \ubeta_j = (\ux_i^{\ast})^{\top} \ubeta_j^{\ast},\label{mean1}
\end{equation}
where $\ux_i^{\ast} = (1,\ux_i^{\top})^{\top}$ and $\ubeta_{j}^{\ast} = (\beta_{0j}, \ubeta_j^{\top})^{\top}$. As we detail below, the linear predictor may include parametric and nonparametric terms. Even though it may appear from (\ref{mean1}) that all regression equations have the same set of predictors, the introduction of binary indicators for variable selection will allow each response to have its own set of covariates.  
 
The implied model for the mean of vector $\uY_i$ is
\begin{equation}
E(\uY_i) = \umu_i = \umu(\uX_i,\ubeta^{\ast}) = \ubeta_0 + \uX_i \ubeta, \nonumber
\end{equation} 
where 
\[
\ubeta_0 =
\begin{pmatrix}
\beta_{01} \\
\beta_{02} \\
\vdots   \\
\beta_{0p} \\
\end{pmatrix},
\uX_i =
\begin{bmatrix}
\ux_i^{\top}  & \uzero^{\top} & \dots & \uzero^{\top} \\
\uzero^{\top} & \ux_i^{\top}  & \dots & \uzero^{\top} \\
\vdots          &                 &       &           \\   
\uzero^{\top} & \uzero^{\top} & \dots & \ux_i^{\top}  \\
\end{bmatrix},
\text{\;and\;}
\ubeta =
\begin{pmatrix}
\ubeta_1 \\
\ubeta_2 \\
\vdots   \\
\ubeta_p \\
\end{pmatrix}.
\]
The mean vector can also be written as $\umu_i = \uX_i^{\ast} \ubeta^{\ast}$, where $\uX_i^{\ast}$ and $\ubeta^{\ast}$ have the same structure as $\uX_i$ and $\ubeta$ above, but with 
$\ux_i$ and $\ubeta_j$ replaced by $\ux_i^{\ast}$ and $\ubeta_j^{\ast}$, $j=1,\dots,p$.   

We let $\uSigma_i$ denote the covariance matrix of the $i$th response vector 
\begin{equation}
\text{cov}(\uY_i)= \uSigma_i = \uSigma(\uR,\uz_i,\ualpha,\usigma^2).\label{covmod1}
\end{equation}
We factorize $\uSigma_i = \uS_i^{1/2} \uR \uS_i^{1/2}$ into a matrix of correlations $\uR$ and a diagonal matrix of variances 
$\uS_{i} = \text{diag}(\sigma^2_{i1},\dots,\sigma^2_{ip})$. The variances $\sigma^2_{ij},j=1,\dots,p,$ are modelled in terms of covariates $\uz_i$ using 
$\sigma^2_{ij} = \sigma^2_{j} \exp(\uz^{\top}_{i} \ualpha_{j})$, where $\ualpha_{j}$ is a vector of regression coefficients and $\sigma^2_{j}$ is a multiplicative variance term. 
Let $\usigma^2 = (\sigma^2_{1},\dots,\sigma^2_{p})^{\top}$ and $\ualpha=(\ualpha^{\top}_1,\dots,\ualpha^{\top}_p)^{\top}$. Clearly, $\uSigma_i$ depends on
$\uR, \uz_i, \ualpha$, and $\usigma^2$, and this is emphasised by the notation in (\ref{covmod1}).

The model specification is completed by assuming a normal distribution for the response vector 
\begin{equation}
\uY_{i} \sim N(\uX_i^{\ast} \ubeta^{\ast}, \uSigma_i), i=1,2,\dots,n. \label{com1} 
\end{equation}
Alternatively, the model can be written in the usual form
\begin{equation}
\uY \sim N(\uX^{\ast} \ubeta^{\ast}, \uSigma), \label{com2} 
\end{equation}
where $\uY=(\uY_1^{\top},\dots,\uY_n^{\top})^{\top}$, 
$\uX^{\ast} = [(\uX_1^{\ast})^{\top},\dots,(\uX_n^{\ast})^{\top}]^{\top}$,  
and $\uSigma = \text{diag}(\uSigma_i, i=1,\dots,n)$.

In the following subsections we detail how the mean and covariance functions are modelled nonparametrically. 

\subsection{Mean model}\label{mean_model}

The mean function $\mu_{ij}=\mu(\ux_i,\ubeta_j^{\ast})$ takes the following general form  
\begin{equation}
\mu_{ij} = \beta_{0j} + \sum_{k=1}^{K_1} u_{ik} \beta_{jk} + \sum_{k=K_1+1}^{K} f_{\mu,j,k}(u_{ik}),\label{mean2}
\end{equation}
where $u_{ik}, k=1,\dots,K_1,$ denotes the regressors with parametrically modelled effects and $u_{ik}, k=K_1+1,\dots,K,$ 
denotes the regressors with effects that are modelled as smooth functions. Further, $K$ denotes the total number of regressors 
that enter the $p$ mean models.

When the assumption of the linearity of the effects of a covariate on the mean function is unrealistic or suspect, it can be relaxed
by the use of smooth functions $f_{\mu,j,k}(.)$, as these can capture non-linear effects. They are represented using 
\begin{equation}
f_{\mu,j,k}(u_{ik}) = \sum_{l=1}^{q_{\mu k}} \beta_{jkl} \phi_{\mu kl}(u_{ik}) = \ux_{ik}^{\top} \ubeta_{jk},\label{fmjk}
\end{equation}
where $\ux_{ik} = (\phi_{\mu k1}(u_{ik}),\phi_{\mu k2}(u_{ik}),\dots,\phi_{\mu kq_{\mu k}}(u_{ik}))^{\top}$ and $\ubeta_{jk} = (\beta_{jk1},\beta_{jk2},\dots,\beta_{jkq_{\mu k}})^{\top}$ 
are the vectors of basis functions and regression coefficients.
In the current article, the basis functions of choice are the radial basis functions, given by 
$\ux_{ik} = \left(u_{ik}, |u_{ik}-\xi_{k1}|^2 \log\left(|u_{ik}-\xi_{k1}|^2\right), \dots, 
|u_{ik}-\xi_{kq_{\mu k}-1}|^2 \log\left(|u_{ik}-\xi_{kq_{\mu k}-1}|^2\right)\right)^{\top}$, where 
$\xi_{k1},\dots,\xi_{kq_{\mu k}-1}$ are the knots. 

Now, model (\ref{mean2}) can be linearised and expressed as model (\ref{mean1}) 
\begin{equation}
\mu_{ij} = \beta_{0j} + \sum_{k=1}^{K_1} u_{ik} \beta_{jk} + \sum_{k=K_1+1}^{K} \ux_{ik}^{\top} \ubeta_{jk} = 
\beta_{0j} + \ux_i^{\top} \ubeta_j,\label{mean3}
\end{equation}
where $\ux_i = (u_{i1},\dots,u_{iK_1},\ux^{\top}_{iK_1+1},\dots,\ux^{\top}_{iK})^{\top}$
and $\ubeta_j=(\beta_{j1},\dots,\beta_{jK_1},\ubeta_{jK_1+1}^{\top},\dots,\ubeta_{jK}^{\top})^{\top}$. 
 
Our general approach for representing smooth functions is to utilize a large number of basis functions. 
With this approach, under-fitting may be avoided. \citet{Chan06} used the same strategy to capture covariate effects 
that are locally adaptive, that is, effects that vary rapidly in some parts of the covariate space and slowly 
in some other parts. We deal with potential over-fitting by allowing positive prior probability that the regression coefficients are exactly zero.     
This is achieved by the introduction of binary variables that allow coefficients to drop out of the model. These, for 
parametric effects, are denoted as $\gamma_{jk} = I[\beta_{jk} \neq 0], k=1,\dots,K_1$, and for nonparametric effects as  
$\gamma_{jkl} = I[\beta_{jkl} \neq 0], k=K_1+1,\dots,K, l=1,\dots,q_{\mu k}$. 
Binary indicators are grouped in the same way as the regression coefficients $\ubeta_j$ after (\ref{mean3}), 
$\ugamma_j=(\gamma_{j1},\dots,\gamma_{jK_1},\ugamma_{j K_1+1}^{\top},\dots,\ugamma_{jK}^{\top})^{\top}$.

Given $\ugamma_j$, model (\ref{mean3}) is expressed as 
\begin{eqnarray}\label{meanfinal}
\mu_{ij} = \beta_{0j} + \ux_{\gamma_j i}^{\top} \ubeta_{\gamma_j j},\nonumber
\end{eqnarray}
where $\ubeta_{\gamma_j j}$ consists of all non-zero elements of $\ubeta_j$
and $\ux_{\gamma_j i}$ of the corresponding elements of $\ux_{i}$. 
Likewise, letting $\ugamma = (\ugamma_1^{\top},\dots,\ugamma_p^{\top})^{\top}$, 
the mean model implied by (\ref{com1}) and (\ref{com2}) may be expressed as 
$E(\uY_i) = \uX_{\gamma i}^{\ast} \ubeta_{\gamma}^{\ast}$ and  
$E(\uY) = \uX_{\gamma}^{\ast} \ubeta_{\gamma}^{\ast}$. 

\subsection{Covariance model}

A first step in modelling the covariance matrices $\uSigma_i$ in terms of covariates is to employ
the separation strategy of \citet{Barnard00}, according to which $\uSigma_{i}$ is expressed as a 
diagonal matrix of variances, $\uS_{i} = \text{diag}(\sigma^2_{i1},\dots,\sigma^2_{ip})$, and a 
correlation matrix $\uR$,   
\begin{equation}
\uSigma_{i} = \uS_{i}^{1/2} \uR \uS_{i}^{1/2}.\label{covmodf}
\end{equation}
The next subsections consider models for the diagonal elements of $\uS_i$ and for the correlation matrix $\uR$. 

\subsubsection{Diagonal variance matrices}

Modelling the diagonal matrices $\uS_{i}$ in terms of covariates is straight forward as the only requirement 
on these elements is that they are nonnegative. Hence, an additive model with a log-link may be utilised
\begin{equation}
\log \sigma^2_{ij} = \alpha_{0j} + \sum_{k=1}^{Q_1} v_{ik} \alpha_{jk} + \sum_{k=Q_1+1}^{Q} f_{\sigma,j,k}(v_{ik}),\label{var1}
\end{equation}
where $v_{ik}, k=1,\dots,Q_1,$ and $v_{ik}, k=Q_1+1,\dots,Q,$ denote covariates with parametric and 
nonparametric effects on the log-variance, respectively. Further, $Q$ denotes the total number of 
effects that enter the $p$ variance models.
Additionally, $f_{\sigma,j,k}(.)$ are smooth functions of covariates, represented as linear combinations of $q_{\sigma k}$ 
radial basis functions and regression coefficients. By analogy to (\ref{fmjk}), we write $f_{\sigma,j,k}(v_{ik}) = \uz_{ik}^{\top} \ualpha_{jk}$.
Hence, by analogy to (\ref{mean3}), model (\ref{var1}) may be written as  
\begin{equation}
\log \sigma^2_{ij} = \alpha_{0j} + \uz^{\top}_{i} \ualpha_{j}, \label{var2}
\end{equation}
where $\uz_i = (v_{i1},\dots,v_{iQ_1},\uz^{\top}_{iQ_1+1},\dots,\uz^{\top}_{iQ})^{\top}$
and $\ualpha_j=(\alpha_{j1},\dots,\alpha_{jQ_1},\ualpha_{jQ_1+1}^{\top},\dots,\ualpha_{jQ}^{\top})^{\top}$.

Consider now vectors of indicator variables for selecting the elements of $\uz_i$ that enter the $j$th variance regression model.
In line with the indicator variables for the mean model, these are denoted by 
$\udelta_j=(\delta_{j1},\dots,\delta_{jQ_1},\udelta_{jQ_1+1}^{\top},\dots,\udelta_{jQ}^{\top})^{\top}$.
Given $\udelta_j$, model (\ref{var2}) becomes
\begin{eqnarray}
\log \sigma^2_{ij} = \alpha_{0j} + \uz^{\top}_{\delta_j i} \ualpha_{\delta_j j},\nonumber
\end{eqnarray}
or equivalently
\begin{eqnarray}
\sigma^2_{ij} = \exp(\alpha_{0j}) \exp(\uz^{\top}_{\delta_j i} \ualpha_{\delta_j j}) =
\sigma^2_{j} \exp(\uz^{\top}_{\delta_j i} \ualpha_{\delta_j j}).\nonumber
\end{eqnarray}

Let $\usigma^2_j = (\sigma_{1j}^2,\dots,\sigma_{nj}^2)^{\top}$. Then, the model for $\usigma^2_j$ can be expressed as 
\begin{equation}\label{mv1}
\usigma^2_j = \sigma^2_{j} \exp(\uZ_{\delta_j} \ualpha_{\delta_j j}), 
\end{equation}
where the design matrix $\uZ_{\delta_j} = [\uz_{\delta_j 1}, \dots,\uz_{\delta_j n}]^{\top}$
consists of $n$ rows, with the $i$th row containing the elements of $\uz_i$ that corresponds to the non-zero elements of $\udelta_j$. 

\subsubsection{Common correlations model}

Turning our attention to the correlation matrix $\uR$, the first prior model we consider, termed the `common correlations
model', takes the following form  
\begin{equation}\label{priorR}
f(\uR|\mu_R,\sigma^2_{R}) = \nu(\mu_{R},\sigma^2_{R}) 
\prod_{k<l} \exp\{-[g(r_{kl})-\mu_{R}]^2/2\sigma^2_{R}\} J[g(r_{kl}) \rightarrow r_{kl}] I[\uR \in \mathcal{C}].
\end{equation}
Here $\mathcal{C}$ denotes the space of correlation matrices, $I[.]$ is the indicator function that ensures the correlation 
matrix is positive definite and $\nu(.,.)$ is the normalizing constant
\begin{eqnarray}\nonumber
\nu^{-1}(\mu_{R},\sigma_{R}^2) = \int_{\uR \in \mathcal{C}} \prod_{k<l} 
\exp\{-[g(r_{kl})-\mu_{R}]^2/2\sigma^2_{R}\} J[g(r_{kl}) \rightarrow r_{kl}] dr_{kl}.
\end{eqnarray}
Function $g(r)$ may be taken to be the Fisher's $z$ transformation $g(r) = \log([1+r]/[1-r])/2$,
considered within Bayesian hierarchical modelling by \citet{DanielsKass99}. 
With this choice, $J[g(r) \rightarrow r] = (1-r)^{-1} (1+r)^{-1}$. Another choice 
is the identity function $g(r)=r$ that simplifies the model formulation.

Making the simplifying model choice of $g(r)=r$ and ignoring the normalizing constant, (\ref{priorR}) reduces to 
\begin{equation}
f(\uR|\mu_R,\sigma^2_{R}) \propto 
\prod_{k<l} \exp\{-(r_{kl}-\mu_{R})^2/2\sigma^2_{R}\} I[\uR \in \mathcal{C}],\label{common1}
\end{equation}
where the product is over the nonredundant, upper triangular, elements of $\uR$ and the kernel is that of a normal density with mean $\mu_R$ and variance $\sigma^2_R$. 
Although it may appear that $\{r_{kl}: k<l\}$ are independent, this is not the case as the indicator function restricts the range of the correlations and induces dependence among them.
The `common correlations model' is intuitive and easy to interpret, however it can can be quite restrictive since all correlations are tied to a common mean $\mu_R$ and a common variance $\sigma^2_R$. For this reason, we consider two models that are more flexible, the `grouped correlations' and `grouped variables' models. 

\subsubsection{Grouped correlations model}

The `grouped correlations model' includes a clustering on the elements of $\uR$, and it takes the form 
\begin{eqnarray}\label{priorR3}
&& f(\uR|\umu_{R},\sigma^2_{R},\ulambda) = \nu(\umu_{R},\sigma^2_{R},\ulambda) \nonumber\\
&& \times \prod_{k<l} \left\{ 
\sum_{h=1}^H I[\lambda_{kl}=h] \exp\{-[g(r_{kl})-\mu_{R,h}]^2/2\sigma^2_{R}\}
\right\} J[g(r_{kl}) \rightarrow r_{kl}] I[\uR \in \mathcal{C}],
\end{eqnarray}
where $H$ denotes the number of correlation groups and $\mu_{R,h}$ denotes the mean of the $h$th group, $h=1,\dots,H$.

Consider, for example, the case depicted in Figure \ref{plotcor}: a correlation matrix of a five-dimensional response, where the $10$ nonredundant correlations are partitioned into $H=3$ groups, namely, $A = \{r_{12}, r_{13}, r_{23}\}$, $B = \{r_{14}, r_{15}, r_{24}, r_{25}, r_{34}, r_{35}\}$, and the singleton group $C=\{r_{45}\}$,
where each group has its own mean. Making the same simplifying choices as those that gave rise to (\ref{common1}), prior (\ref{priorR3}) for the current scenario can be written as
\begin{eqnarray}
&&f(\uR|\umu_{R},\sigma^2_{R},\ulambda) \propto  
\prod_{r_{kl} \in A} \exp\{-(r_{kl}-\mu_{R,A})^2/2\sigma^2_{R}\} \nonumber\\
&&\times \prod_{r_{kl} \in B} \exp\{-(r_{kl}-\mu_{R,B})^2/2\sigma^2_{R}\}
\prod_{r_{kl} \in C} \exp\{-(r_{kl}-\mu_{R,C})^2/2\sigma^2_{R}\} \nonumber
I[\uR \in \mathcal{C}]
\end{eqnarray}

\subsubsection{Grouped variables model}\label{gv}

The `grouped variables model' is another clustering model that clusters the variables instead of the correlations.
The prior takes the form
\begin{eqnarray}\label{priorR4}
&& f(\uR|\umu_{R},\sigma^2_{R},\ulambda) = \nu(\umu_{R},\sigma^2_{R},\ulambda) \nonumber\\ 
&& \times \prod_{k<l} \left\{ \sum_{h_1, h_2=1}^G I[\lambda_{k}=h_1] I[\lambda_{l}=h_2] 
\exp\{-[g(r_{kl})-\mu_{R,h_1,h_2}]^2/2\sigma^2_{R}\} \right\} J[g(r_{kl}) \rightarrow r_{kl}]
I[\uR \in \mathcal{C}],\nonumber
\end{eqnarray}
where $G$ is the number of groups in which the variables are partitioned, creating $H=G(G+1)/2$ clusters for the correlations.  

\begin{figure}
\begin{center}
\includegraphics[width=0.3\textwidth]{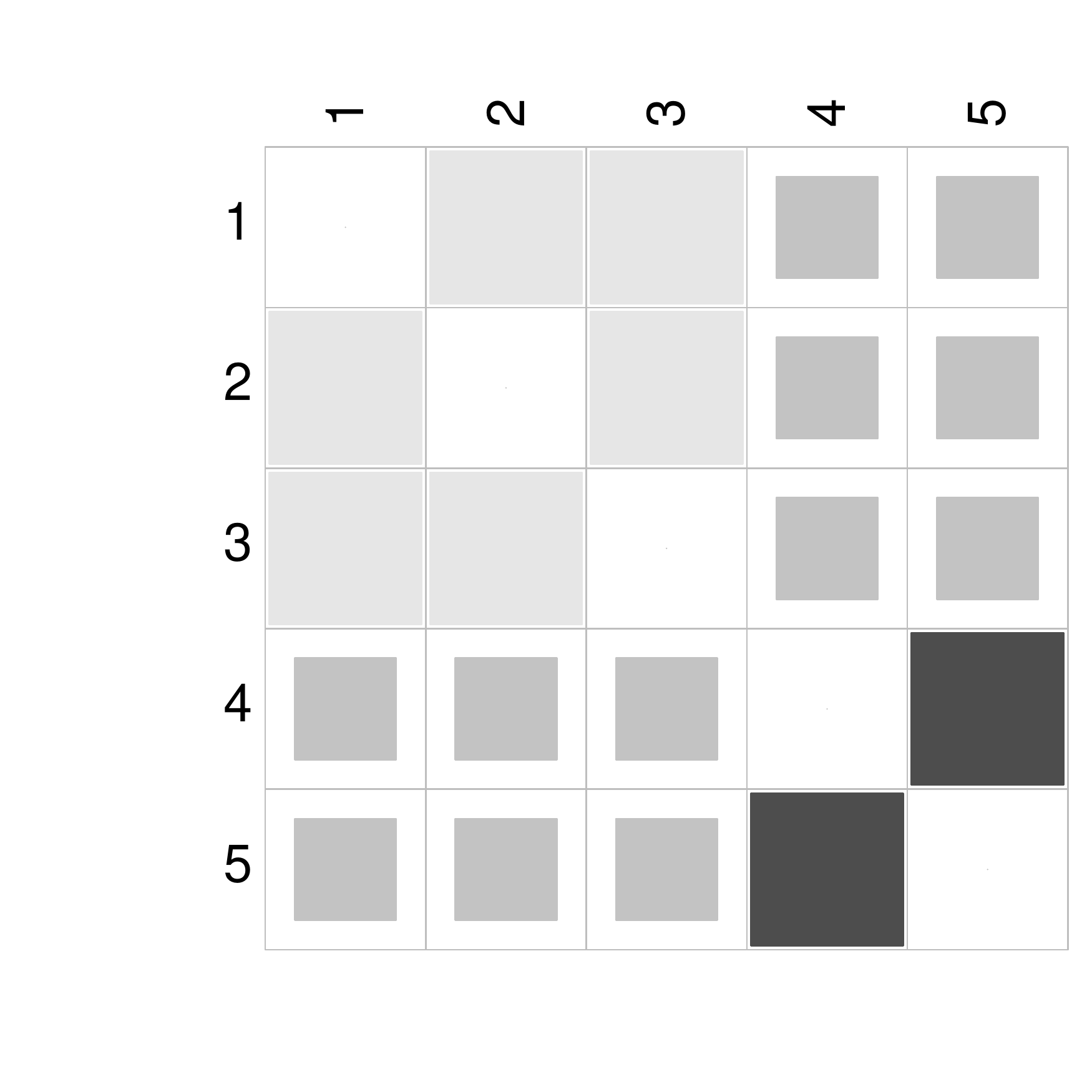}
\end{center}
\caption{A $5 \times 5$ correlation matrix with two groups of variables, $\{1,2,3\}$ and $\{4,5\}$, and three 
groups of correlations, denoted by different colours.}\label{plotcor}
\end{figure}

A clustering on the variables is more structured than a clustering on the correlations. In other words, a clustering 
on the variables implies a clustering on the correlations. The converse, however, is not necessarily true.  
Revisiting Figure \ref{plotcor}, we see that the $p=5$ responses are grouped into two clusters, the first group consisting of variables $\{1, 2, 3\}$,
and the second one of variables $\{4, 5\}$. These two groups create three groups of correlations, two of which
describe the correlations within each group and one that describes the correlation between the two groups.  

\subsection{Prior specification}\label{prior_spec}

Let $\utX = \uSigma^{-\frac{1}{2}} \uX^{\ast}$. 
The prior for $\ubeta_{\gamma}^{\ast}$ is specified as \citep{AZ}
\begin{eqnarray}
\ubeta_{\gamma}^{\ast} | c_{\beta}, \ugamma, \ualpha, \udelta, \uR \sim N(\uzero,c_{\beta} (\utX_{\gamma}^{\top} \utX_{\gamma} )^{-1}).\label{gprior}
\end{eqnarray}
Further, the prior for $c_{\beta}$ is specified as inverse Gamma, $c_{\beta} \sim \text{IG}(a_{\beta},b_{\beta})$.

For the vector $\ugamma_j=(\gamma_{j1},\dots,\gamma_{jK_1},\ugamma_{j K_1+1}^{\top},\dots,\ugamma_{jK}^{\top})^{\top}, j=1,\dots,p,$ of indicator variables, we specify independent binomial priors for each of its $K$ subvectors, 
\begin{eqnarray}
P(\ugamma_{jk}|\pi_{\mu j k}) = \pi_{\mu j k}^{N(\gamma_{jk})} (1-\pi_{\mu j k})^{q_{\mu k}-N(\gamma_{jk})},  k=1,\dots,K,\nonumber
\end{eqnarray}
where $N(\gamma_{jk}) = \gamma_{jk}$ for parametric effects, $k=1,\dots,K_1$, and $N(\gamma_{jk}) = \sum_{l=1}^{q_{\mu k}} \gamma_{jkl}$
for nonparametric effects, $k=K_1+1,\dots,K.$
We work with Beta priors for $\pi_{\mu j k}$,
$\pi_{\mu j k} \sim \text{Beta}(c_{\mu j k},d_{\mu j k})$, $j=1,\dots,p, k=1,\dots,K$,
although sparsity inducing, zero-inflated Beta priors, are also an attractive option. 

Continuing with the priors on the covariance parameters, we specify independent normal priors for $\ualpha_{\delta_j j}$
\begin{eqnarray}
\ualpha_{\delta_j j} | c_{\alpha j}, \udelta_j \sim N(\uzero,c_{\alpha j} \uI), j=1,\dots,p.\nonumber
\end{eqnarray}
Further,  the priors we consider for $c_{\alpha j}$, are the half-normal, 
$\sqrt{c_{\alpha j}} \sim \text{HN}(\phi^2_{c_{\alpha j}}) \equiv N(0,\phi^2_{c_{\alpha j}}) I[\sqrt{c_{\alpha j}}>0]$, 
and the inverse Gamma, $c_{\alpha j} \sim \text{IG}(a_{\alpha j},b_{\alpha j}),$ 
$j=1,\dots,p.$

For the $Q$ subvectors of $\udelta_{j}=(\delta_{j1},\dots,\delta_{jQ_1},\udelta_{jQ_1+1}^{\top},\dots,\udelta_{jQ}^{\top})^{\top}, j=1,\dots,p,$ we specify independent binomial priors 
\begin{eqnarray}
P(\udelta_{jk}|\pi_{\sigma j k}) = \pi_{\sigma j k}^{N(\delta_{jk})} (1-\pi_{\sigma j k})^{q_{\sigma k}-N(\delta_{jk})},k=1,\dots,Q,\nonumber
\end{eqnarray}
where $N(\delta_{jk})=\delta_{jk}$ for parametric effects, $k=1,\dots,Q_1$, and $N(\delta_{jk}) = \sum_{k=1}^{q_{\sigma k}} \delta_{jkl}$
for nonparametric effects, $k=Q_1+1,\dots,Q.$
We specify independent Beta priors for $\pi_{\sigma j k},$ $\pi_{\sigma j k} \sim \text{Beta}(c_{\sigma j k},d_{\sigma j k})$, $j=1,\dots,p, k=1,\dots,Q$.

For $\sigma^2_{j}, j=1,\dots,p,$ we consider inverse Gamma and half-normal priors, denoted as 
$\sigma^2_{j} \sim \text{IG}(a_{\sigma j},b_{\sigma j})$ and $\sigma_{j} \sim \text{HN}(\phi^2_{\sigma j}) \equiv N(0,\phi^2_{\sigma j}) I[\sigma_{j} > 0]$.

Lastly, we describe the priors on the parameters of the correlation models.   
Starting with the `common correlations model' in (\ref{priorR}), we place the following priors 
on its parameters
\begin{equation}
\mu_R \sim N(0,\varphi^2_R) \text{\;and\;} \sigma_R \sim \text{HN}(\phi^2_R) \equiv N(0,\phi^2_R) I[\sigma_R>0]. \nonumber
\end{equation} 

We take the `grouped correlations model' to be arising from the `common correlations model', by treating the 
prior on $\mu_R$ as another unknown model parameter. In symbols, $\mu_R \sim P$, where $P$ is an unknown distribution.  
Here, we place a Dirichlet process (DP) prior on $P$ \citep{Fer73}.
Due to the almost sure discreteness of the DP, the prior $P$ admits the following representation
\begin{equation}
P(.) = \sum_{h=1}^{\infty} w_h \delta_{\mu_{R,h}}(.),\nonumber
\end{equation}
where $\delta_x(.)$ is an indicator function, $\delta_x(y) = I[x=y]$. 
The prior weights $w_h$ are constructed utilising the so called stick-breaking process \citep{sethuraman}.
Let $v_h, h=1,2,\dots,$ be independent draws from a $\text{Beta}(1,\alpha^{*})$ distribution. We have, 
$w_1 = v_1$, for $ l \geq 2$, $w_l = v_l \prod_{h=1}^{l-1} (1-v_h)$.  We take the concentration parameter $\alpha^{*}$ to be unknown
and we assign to it a gamma prior $\alpha^{*} \sim \text{Gamma}(a_{\alpha*},b_{\alpha*})$ with mean $a_{\alpha*}/b_{\alpha*}$.
Further, $\mu_{R,h}$ are generated from the so called base distribution, here taken to be $N(0,\varphi^2_R)$.

The `grouped correlations model' in (\ref{priorR3}) is obtained by first writing
\begin{eqnarray}
&&\int_{\mu_R} f(\uR|\mu_R,\sigma^2_{R}) dP(\mu_R) =
\sum_{h=1}^{\infty} w_h  f(\uR|\mu_{R,h},\sigma^2_{R}) = \nonumber\\
&& \nu(\umu_{R},\sigma^2_{R},\uw) 
\sum_{h=1}^{\infty} w_h \prod_{k<l} 
\exp\{-[g(r_{kl})-\mu_{R,h}]^2/2\sigma^2_{R}\}
J[g(r_{kl}) \rightarrow r_{kl}] I[\uR \in \mathcal{C}],\nonumber
\end{eqnarray}
where $\umu_R$ and $\uw$ denote the vectors of group means and the stick-breaking weights, respectively. 
In practice, we truncate $P()$ to include $H$ components. In this case, the prior weights are constructed
as before, except for the $H$th one that is now constructed as $w_H = \prod_{h=1}^{H-1} (1-v_h)$. 
Further, we introduce allocation variables $\lambda_{kl}$ to indicate the component in which 
$r_{kl}$ is assigned to, $k=1,\dots,p, k < l$. The stick-breaking weights provide the prior on the
allocation variables: $P(\lambda_{kl} = h) = w_h, h=1,\dots,H$. With these observations, it is clear how model 
(\ref{priorR3}) follows. 

The development on the `grouped variables model' is very similar, with the clustering now performed on the 
variables rather than the correlations. 

In the simulation study and applications that we present in Sections \ref{sim} and \ref{applications}, 
we use the following priors, unless otherwise stated within the relevant sections. 
For $c_{\beta},$ we specify 
$\text{IG}(1/2, np/2),$ as a $p$-variate analogue of the prior of \citet{liang_mixtures_2008}. For all inclusion probabilities, $\pi_{\mu j k}$ and $\pi_{\sigma j k}$, we define Beta$(1,1)$, that is uniform priors. 
The prior on all $c_{\alpha j}$ is specified to be IG$(1.1,1.1)$. Further, for all $\sigma_j$, we define the prior to be HN$(2)$. In addition, we specify $\mu_R \sim N(0,1)$ and $\sigma_R \sim \text{HN}(1)$. Lastly, the DP base distribution is taken to be the standard normal while the concentration is taken to have a $\alpha^{\ast} \sim \text{Gamma}(5,2)$ prior.

\section{Posterior Sampling}\label{postsamp}

To carry out posterior sampling we consider two likelihood functions and use the one that is more computationally convenient for each 
step of the MCMC algorithm. 

We first consider the full likelihood, that is, the one that involves all model parameters. The contribution of $\uY_i, i=1,\dots,n$, using decomposition (\ref{covmodf}), may be expressed as 
\begin{eqnarray}
&& f(\uY_i|\ubeta^{\ast},\ugamma,c_{\beta},\ualpha,\udelta,\usigma^2,\uR) \propto  
|\uSigma_i(\uR,\ualpha,\udelta,\usigma^2)|^{-\frac{1}{2}} 
\exp\{-(\uY_i - \uX_i^{\ast} \ubeta^{\ast})^{\top} \uSigma_i^{-1}(\uY_i - \uX_i^{\ast} \ubeta^{\ast})/2\}  \nonumber\\
&&\propto
|\uS_i^{\frac{1}{2}} \uR \uS_i^{\frac{1}{2}} |^{-\frac{1}{2}} 
\exp\{- (\uS_i^{-\frac{1}{2}} \ur_i)^{\top} \uR ^{-1} (\uS_i^{-\frac{1}{2}} \ur_i)/2\}
\propto
|\uS_i|^{-\frac{1}{2}} 
|\uR|^{-\frac{1}{2}}
\exp\{-\text{tr}(\uR ^{-1} \utS_i)/2\},\nonumber
\end{eqnarray}
where $\ur_i = \uY_i - \uX_i^{\ast} \ubeta^{\ast}$ and 
$\utS_i = (\uS_i^{-1/2} \ur_i) (\uS_i^{-1/2} \ur_i)^{\top}$. Hence, the likelihood function, based on all observations, is 
\begin{eqnarray}\label{full}
f(\uY|\ubeta^{\ast},\ugamma,c_{\beta},\ualpha,\udelta,\usigma^2,\uR) \propto  
\prod_{i=1}^n|\uS_i|^{-\frac{1}{2}} 
|\uR|^{-\frac{n}{2}}
\exp\{-\text{tr}(\uR ^{-1} \utS_i)/2\}
\end{eqnarray}
where $\utS = \sum_{i=1}^n \utS_i$.

To improve mixing of the MCMC algorithm, we can integrate out vector $\ubeta^{\ast}$ from the likelihood (\ref{full}), to obtain
\begin{eqnarray}\label{marginal}
f(\uY|\ugamma,c_{\beta},\ualpha,\udelta,\usigma^2,\uR) = (2\pi)^{-\frac{np}{2}} 
|\uSigma(\uR,\ualpha,\udelta,\usigma^2)|^{-\frac{1}{2}} (c_{\beta}+1)^{-\frac{N(\gamma)+p}{2}} \exp(-S/2),
\end{eqnarray}
where
\begin{eqnarray}
S = S(\uY,\ugamma,c_{\beta},\ualpha,\udelta,\usigma^2,\uR) = \utY^{\top} \big( I - 
\frac{c_{\beta}}{1+c_{\beta}} \utX_{\gamma} (\utX_{\gamma}^{\top}\utX_{\gamma})^{-1} \utX_{\gamma}^{\top} \big) \utY, \nonumber
\end{eqnarray}
with $\utY =  \uSigma^{-\frac{1}{2}} \uY$ and $N(\gamma)+p$ the total number of columns in $\utX_{\gamma}$. 
We compute $S$ using the following, more convenient, expression  
\begin{eqnarray} 
S = \text{tr}\big(\uR^{-1} \sum_{i=1}^n \ucy_{i} \ucy_{i}^{\top}\big)  -
c_{\beta} (1+c_{\beta})^{-1} \big(\sum_{i=1}^n \ucy_{i}^{\top} \uR^{-1} \ucX_{\gamma i}\big) 
\big(\sum_{i=1}^n \ucX_{\gamma i}^{\top} \uR^{-1} \ucX_{\gamma i}\big)^{-1}
\big(\sum_{i=1}^n \ucX_{\gamma i}^{\top} \uR^{-1} \ucy_{i}\big),\nonumber
\end{eqnarray}
where $\ucX_{\gamma i} = \uS_i^{-1/2} \uX_{\gamma i}^{\ast}$ and $\ucy_{i} = \uS_i^{-1/2} \uY_{i}$.

Sampling from the posterior of the parameters of the correlation matrices poses the greatest challenge.  
Consider, for instance, sampling from the posterior of parameter $\mu_R$ of the `common 
correlations model', given in (\ref{priorR}), using the Metropolis-Hastings algorithm. Letting $\mu_R^C$ and $\mu_R^P$ denote current and 
proposed values, the acceptance probability will involve the ratio of the normalising constants 
$\nu(\mu_{R}^P,\sigma_{R}^2)/\nu(\mu_{R}^C,\sigma_{R}^2)$, which can be very computationally demanding
to calculate.  
Posterior sampling, however, may be simplified by utilising the `shadow prior' \citep{Liechty}. 
The basic idea is to introduce latent variables $\theta_{kl}$ between the correlations $r_{kl}$ and the mean $\mu_R$, 
by which prior (\ref{priorR}) becomes 
\begin{equation}\label{priorR2}
f(\uR|\utheta,\tau^2) = \nu(\utheta,\tau^2) 
\prod_{k<l} \exp\{-[g(r_{kl})-\theta_{kl}]^2/2\tau^2\} J[g(r_{kl}) \rightarrow r_{kl}] I[\uR \in \mathcal{C}],
\end{equation}
where 
\begin{eqnarray}\nonumber
\nu^{-1}(\utheta,\tau^2) = \int_{\uR \in \mathcal{C}} \prod_{k<l} 
\exp\{-[g(r_{kl})-\theta_{kl}]^2/2\tau^2\} J[g(r_{kl}) \rightarrow r_{kl}] dr_{kl}.
\end{eqnarray}
Further, variables $\theta_{kl}$ are assumed to be independently distributed as 
\begin{equation}
\theta_{kl} \sim N(\mu_R,\sigma^2_R),l=1,\dots,p, k<l,\label{thetaP}
\end{equation}
and $\tau$ is taken to be a small constant. 
Sampling from the posterior of $\utheta=\{\theta_{kl}\}$ still involves the ratio of the normalising constants, 
$\nu(\utheta^P,\tau^2)/\nu(\utheta^C,\tau^2)$, but that, as was argued by \citet{Liechty}, for small $\tau$, can reasonably be approximated by one. 
In addition, now sampling for the posterior of $\mu_R$ given $\utheta$ is straight forward. 
Hence, the computational burden is greatly alleviated. 

We now provide details on the step of the MCMC algorithm that updates $\uR$. This step uses the prior in (\ref{priorR2}) and the likelihood in (\ref{full}). 
Hence, the posterior of $\uR$ is
\begin{eqnarray}\label{postRt}
f(\uR|\dots) \propto |\uR|^{-\frac{n}{2}} \exp\{- \text{tr}(\uR^{-1}\utS)/2\} 
\prod_{k<l} \exp\{-[g(r_{kl})-\theta_{kl}]^2/2\tau^2\} J[g(r_{kl}) \rightarrow r_{kl}] I[\uR \in \mathcal{C}]. \label{Rt}
\end{eqnarray}

To obtain a proposal density and sample from (\ref{Rt}) we utilize the method of \citet{xiao} and \citet{LD2006}.
We start by considering a symmetric, positive definite and otherwise unconstrained matrix $\uE$ in place of $\uR$,
assumed to have an inverse Wishart prior $\uE \sim \text{IW}(\zeta,\uPsi)$, 
with mean equal to the realization of $\uE$ from the previous iteration of the sampler.  
Given the inverse Wishart prior on $\uE$, we obtain the following, easy to sample from, inverse Wishart posterior
\begin{eqnarray}
g(\uE|\dots) \propto |\uPsi|^{\frac{\zeta}{2}} |\uE|^{-\frac{n+\zeta+p+1}{2}} \exp\{-\text{tr}[\uE^{-1} (\utS+\uPsi)]/2\}.\label{Et}
\end{eqnarray}
We decompose $\uE=\uD^{1/2} \uR \uD^{1/2}$ into a diagonal matrix of variances  
$\uD = \text{diag}(d^2_{1},\dots,d^2_{p})$, 
and a correlation matrix $\uR$. The Jacobian associated with this transformation is 
$J(\uE \rightarrow \uD, \uR) = \prod_{k=1}^p (d_{k})^{p-1} = |\uD|^{(p-1)/2}$. 
It follows that the joint density for $(\uD, \uR)$ is  
\begin{eqnarray}
h(\uD, \uR|\dots) \propto |\uPsi|^{\frac{\zeta}{2}} |\uD|^{(p-1)/2} |\uE|^{-\frac{n+\zeta+p+1}{2}} \exp\{-\text{tr}[\uE^{-1} (\uS+\uPsi)]/2\}.\label{DR}
\end{eqnarray}

Sampling from (\ref{DR}) at iteration $u+1$ proceeds by sampling $\uE^{(u+1)}$ from (\ref{Et}) and decomposing 
$\uE^{(u+1)}$ into $(\uD^{(u+1)},\uR^{(u+1)})$. 
Further, the pair $(\uD^{(u+1)},\uR^{(u+1)})$ is accepted as a sample from (\ref{Rt}) with probability
\begin{eqnarray}
\alpha = \min\left\{1,\frac{f(\uR^{(u+1)}|\dots) h(\uD^{(u)}, \uR^{(u)}|\dots)}
{f(\uR^{(u)}|\dots) h(\uD^{(u+1)}, \uR^{(u+1)}|\dots)}\right\},\nonumber
\end{eqnarray}
where, in $h(,|)$, $\uPsi = (\zeta-p-1) \uE^{(u)}$.
We treat $\zeta$ as a tuning parameter and we automatically adjust its value \citep{roberts_examples_2009} so as to obtain an acceptance probability of 
$20\% - 25\%$ \citep{Roberts2001c}.
Further details on the MCMC steps are provided in the Appendix and the supplementary materials online. 

\section{Simulation study}\label{sim}

The first purpose in this simulation study is to quantify, in a simple scenario, the gains that one may have, in terms of reduced bias and variance, when estimating a posterior mean function by fitting the multivariate model of the highest available response dimension instead of a lower dimensional model. The second one is to report the run-times needed to fit models of increasing response dimension. To achieve these goals, it suffices to consider data-generating mechanisms with simple mean and variance functions. Simulation studies that illustrate the performance of the univariate version of the current model in capturing complex mean and variance functions have been presented by \citet{Chan06} and \citet{pap18}, and hence will not be revisited here.   
Additionally, we evaluate the model's ability to select important variables with its spike-slab priors, and whether the variable selection ability depends on the dimension of the response. Lastly, we compare the performance of the models presented here with the performance of other models for sparse multivariate regression that have appeared in the literature. 

The data-generating mechanism that we consider consists of ten orthogonal covariates, $x_1, \dots, x_{10}$, each generated from a uniform distribution in the $(-0.5,0.5)$ interval, and ten responses, $Y_1, \dots, Y_{10}$, that are  generated from a multivariate normal distribution with mean $\umu = (\beta_{01} + \beta_{11} x_{1},0,\dots,0)^{\top} $ and covariance $ \uSigma(\rho)$.  
The first element of the mean is a linear function of $x_1$, while all other elements are zero. The covariance matrix is taken to 
have diagonal elements equal to one and all off diagonal elements equal to $\rho$.  

The main interest here is on the quality of the estimate of the mean function of the first dimension. We examine how this quality, as measured by the posterior bias and variance, depends on the dimension of the response, the value of the correlation coefficient $\rho$, and the chosen linear predictor. The effect of the dimension of the response is evaluated by fitting one-, two-, four-, six-, and ten-dimensional response models to the dataset that includes the ten responses. The effect of the correlation coefficient is examined by letting $\rho$ take values in the set $\{0.1, 0.3, 0.5, 0.7, 0.9\}$. The effect of the choice of the linear predictor is evaluated by fitting mean models that include only the relevant covariate, $x_1$, models that include the first three covariates, $x_1, x_2, x_3$,  and models that include all available covariates, $x_1,\dots, x_{10}$.

For example, a four-dimensional response model considers $Y_1,\dots,Y_4$ and ignores $Y_5,\dots, Y_{10}$.
The mean functions of the responses are modelled using one of the following three options
\begin{equation}\label{threemm}
\mu_{j} = \beta_{0j} + \beta_{j1} x_{1}, 
\mu_{j} = \beta_{0j} + \sum_{k=1}^3 \beta_{jk} x_{k}, \mu_{j} = \beta_{0j} + \sum_{k=1}^{10} \beta_{jk} x_{k},
\end{equation}
$j=1,2,3,4$, where the first specification is correct for the first response and wrong for the other three responses, while the second and third are wrong for all responses. Further, we fit models with constant variance functions $\sigma^2_j$, $j=1,2,3,4,$ and the common correlations model given in (\ref{priorR}). Both the variance and correlation model specifications are the correct ones. 

The regression coefficients are taken to be $\beta_{01}=0$ and $\beta_{11}=3.47$. The chosen value of $\beta_{11}$ achieves a signal-to-noise ratio (SNR) equal to one, where SNR is defined as $\text{SNR} = (\text{SST} - \text{SSE}) / \text{SSE}$, with SST the total sum of squares SST=$\sum_{i=1}^n(y_{i1}-\bar{y}_1)^2$ and SSE the error sum of squares SSE=$\sum_{i=1}^n(y_{i1}-\hat{y}_{i1})^2$. In addition, two values for the sample size are considered, $n=50, 150$. 

For all models we run the MCMC sampler for $40,000$ sweeps, discarding the first $20,000$ as burn in, and of the remaining $20,000$ keeping one in two. This results in $10,000$ samples for $\mu_{i1} = E(Y_{i1}|x_{i1}) = \beta_{01} + \beta_{11} x_{i1}$, obtained by replacing the regression coefficients in the chosen linear predictor by the corresponding sampled values.  
We recall that the choices of the linear predictor are given in (\ref{threemm}).
Our final estimate of $\mu_{i1}, i=1,\dots,n,$ is taken to be the median of the sampled values, which we denote by $\hat \mu_{i1d}$, where subscript $d$ denotes the dimension of the response, $d=1,2,4,6,10$. We quantify uncertainty about these estimates by forming $90\%$ credible intervals $(\hat \mu_{q_1,i1d}, \hat \mu_{q_2,i1d})$ where the end-points of these intervals are the $5\%$ and $95\%$ quantiles of the sampled values. 

We compare the models in terms of their bias and variance in estimating $\mu_{i1}$. As we estimate $\mu_{i1}$ for a range of $x_1$ values, we summarize the bias by computing the sum of squared deviations of the estimates from the targets, $B(d) = \sum_{i=1}^n(\mu_{1i}-\hat\mu_{i1d})^2$. 
Further, the variance of the estimates is summarized by computing the sum of the squared lengths of the credible intervals, $V(d)=\sum_{i=1}^n(\hat\mu_{q_1,i1d}-\hat\mu_{q_2,i1d})^2$.
To obtain representative results and independent of the generated dataset, we repeat the above process on $40$ replicate datasets
for each sample size $n$ by correlation $\rho$ combination.

Results for the first choice of the mean model, $\mu_{j} = \beta_{0j} + \beta_{j1} x_{1}$, are presented in Tables \ref{bias.sim} and \ref{var.sim}. Table \ref{bias.sim} compares models by 
reporting the ratio $B(d)/B(1)$ (as a percentage), that we refer to as the relative bias, while Table \ref{var.sim} compares 
models by reporting the ratio $V(d)/V(1)$, that we refer to as the relative variance, $d=2,4,6,10$.  
In Table \ref{bias.sim} we see a clear decreasing trend of the relative bias as the correlation between the
responses increases. 
Although the gains are low when the correlation between the responses is low, 
we observe a rapid decrease in the relative bias as the correlation increases, for all sample sizes $n$
and for all $d$. We also observe that for $d=4$, relative bias
is lower than for $d=2$, especially for $n=50$ and for correlations higher than $0.1$. However, relative bias 
for $d=6$ and $d=10$ is very similar to that for $d=4$.
Similar patterns are observed for the relative variances in Table \ref{var.sim}. 
There is a clear decreasing trend as the correlation increases, for all sample sizes and all dimensions. 
This decrease is more pronounced for high correlations between the responses, as one would expect given the results of \citet{Zellner62}. 
Results for the second and third mean models, as displayed in (\ref{threemm}), are available in the supplement.
Generally, the patterns of relative bias and variance are the same as those seen above, however, the gains are generally more pronounced.

It is always useful to compare new methods, such as the one presented here, with methods that have appeared in the literature. Here we make comparisons, in terms of posterior bias, with the method for multivariate regression of \citet{Rothman},
that has been implemented in the R package MRCE \citep{MRCE}. \citet{Rothman} present a method for sparse multivariate regression 
with covariance estimation (henceforth abbreviated as MRCE) that estimates regression coefficients by maximizing a multivariate normal likelihood with lasso 
penalties for the regression coefficients and the elements of the precision matrix. 
To each of the simulated datasets we fit MRCE models of response dimension $d=2,4,6,10$ and compute the total bias, 
$B_M(d) = \sum_{i=1}^n(\mu_{1i}-\tilde\mu_{i1d})^2$, where 
$\tilde \mu_{i1d}$ is either $\tilde \mu_{i1d} = \tilde \beta_{01d} + \sum_{k=1}^3 \tilde \beta_{1kd} x_{ik}$ or 
$\tilde \mu_{i1d} = \tilde \beta_{01d} + \sum_{k=1}^{10} \tilde \beta_{1kd} x_{ik}$, depending on the mean model choice, and
$\tilde \beta_{01d}, \dots, \tilde \beta_{1,3d}, \dots, \tilde \beta_{1,10d}$ are the MRCE coefficient estimates
for the second and third mean models. 
We note that comparisons are based on the second and third the mean models, but not the first one, as 
both approaches have a mechanism for inducing sparseness. 
Results, in the form of ratios $B(d)/B_M(d), d=2,4,6,10,$ expressed as percentages, for the second mean model, 
the two sample sizes and the five correlation coefficients
are presented in Table \ref{comp.sim}. We see that all entries are well below $100\%$, with the minimum at $42.29\%$ and the maximum
at $72.28\%$. Results for the third mean model are available in the supplement, and they are generally more pronounced than those of
 Table \ref{comp.sim}, ranging from $27.44\%$ to $61.39\%$. A major advantage, however, of the MRCE approach is that the resulting algorithm
is less computationally intensive than the MCMC sampler and thus model fitting is typically very fast.

\begin{table}
\begin{center}
\caption{Simulation study results: the entries of the table are the relative biases $B(d)/B(1), d=2,4,6,10$, 
	expressed as percentages. 
Rows refer to the sample size $n = 50, 150,$ and columns to the correlation between the responses, $\rho=
0.1,0.3,0.5,0.7,0.9$. Results are based on the first mean model, and $40$ replicate datasets per sample size by correlation combination.} \label{bias.sim}
\begin{tabular}{l|rrrrr|l|rrrrr}
\hline
$d=2$  &  0.1   &    0.3  &   0.5  &   0.7  &   0.9 & $d=4$  &  0.1   &    0.3  &   0.5  &   0.7  &   0.9\\
\hline
50  &  94.41 & 89.19 & 82.81 & 69.72 & 53.22 & 50  & 93.83 & 81.73 & 69.97 & 59.49 & 50.17\\
150 &  99.42 & 96.71 & 93.41 & 88.14 & 79.25 & 150 & 99.35 & 97.36 & 92.99 & 86.91 & 78.25\\
\hline
$d=6$  &  0.1   &    0.3  &   0.5  &   0.7  &   0.9 & $d=10$  &  0.1   &    0.3  &   0.5  &   0.7  &   0.9\\
\hline
50  &  97.82 & 83.28 & 69.96 & 58.47 & 49.67 & 50  & 103.06 & 81.67 & 70.86 & 58.08 & 49.73\\
150 &  99.41 & 97.00 & 92.01 & 85.33 & 77.17 & 150 &  97.49 & 96.12 & 91.39 & 85.30 & 77.36\\
\hline 
\end{tabular}
\end{center}
\end{table}

\begin{table}
\begin{center}
\caption{Simulation study results: the entries of the table are the relative variances $V(d)/V(1), d=2,4,6,10$, 
	expressed as percentages. 
Rows refer to the sample size $n = 50,150,$ and columns to the correlation between the responses, $\rho=
0.1,0.3,0.5,0.7,0.9$. Results are based on the first mean model, and $40$ replicate datasets per sample size by correlation combination.} \label{var.sim}
\begin{tabular}{l|rrrrr|l|rrrrr}
\hline
$d=2$  &  0.1   &    0.3  &   0.5  &   0.7  &   0.9 & $d=4$  &  0.1   &    0.3  &   0.5  &   0.7  &   0.9\\
\hline
50  & 101.06 & 98.32 & 92.09 & 82.99 & 68.22  & 50  & 100.55 & 94.00 & 85.39 & 74.03 & 60.29\\
150 & 102.15 & 97.77 & 90.27 & 78.80 & 62.66  & 150 & 100.62 & 93.61 & 83.91 & 72.03 & 58.36\\
\hline
$d=6$  &  0.1   &    0.3  &   0.5  &   0.7  &   0.9 & $d=10$  &  0.1   &    0.3  &   0.5  &   0.7  &   0.9\\
\hline
50  & 97.93 & 91.59 & 82.38 & 75.41 & 65.03  & 50  & 95.99 & 88.28 & 81.78 & 73.69 & 65.30\\
150 & 100.34 & 92.14 & 82.34 & 70.71 & 58.14  & 150 & 98.32 & 89.23 & 79.15 & 68.92 & 59.04\\
\hline
\end{tabular}
\end{center}
\end{table}

\begin{table}
	\begin{center}
		\caption{Simulation study results: the entries of the table are the relative biases $B(d)/B_M(d), d=2,4,6,10$, 
			expressed as percentages. 
			Rows refer to the sample size $n = 50, 150,$ and columns to the correlation between the responses, $\rho=
			0.1,0.3,0.5,0.7,0.9$. Results are based on the second mean model, and $40$ replicate datasets per sample size by correlation combination.} \label{comp.sim}
		\begin{tabular}{l|rrrrr|l|rrrrr}
			\hline
			$d=2$  &  0.1   &    0.3  &   0.5  &   0.7  &   0.9 & $d=4$  &  0.1   &    0.3  &   0.5  &   0.7  &   0.9\\
			\hline
			50  & 66.22 & 65.19 & 58.12 & 53.17 & 49.68 & 50  & 61.74 & 55.34 & 47.99 & 50.15 & 53.66\\
			150 & 43.24 & 45.74 & 50.00 & 52.91 & 54.16 & 150 & 42.29 & 49.70 & 53.00 & 53.50 & 61.98\\
			\hline
			$d=6$  &  0.1   &    0.3  &   0.5  &   0.7  &   0.9 & $d=10$  &  0.1   &    0.3  &   0.5  &   0.7  &   0.9\\
			\hline
			50  & 53.46 & 51.28 & 55.13 & 57.24 & 60.15 & 50  & 45.62 & 45.44 & 46.41 & 50.38 & 59.00\\
			150 & 42.47 & 47.91 & 49.78 & 51.67 & 62.42 & 150 & 42.96 & 49.09 & 46.92 & 53.02 & 72.28\\
			\hline 
		\end{tabular}
	\end{center}
\end{table}

To evaluate the variable selection performance of the proposed model, and to check its possible dependence on the response dimension, the correlation coefficient, the sample size, and the mean model choice, we compute the  posterior probabilities that at least one of the irrelevant regressors, $x_2, x_3,$ or $x_2, \dots, x_{10},$ is included in the mean  model of the first response. Results, for the second mean model choice, expressed as percentages, are displayed in Table \ref{vs.sim}. We note that this evaluation depends on the choice of the prior for the inclusion probabilities, described in Section \ref{prior_spec}. The current evaluation is based on a Beta$(1,3)$ prior, but the inclusion probabilities can be made smaller by changing the parameters of the Beta prior in a way that decreases the prior probability of inclusion. We further note that the relevant predictor, $x_1,$ was almost always included in the model, regardless of the choice of the Beta prior. For this reason, we do not provide results on the inclusion probabilities of this regressor. When fitting one-dimensional models, the irrelevant regressors were included $8.26\%$ of the time when $n=50$, and $5.17\%$ of the time when $n=150$. 
From this observation and from Table \ref{vs.sim}, it is clear that the probabilities of inclusion decrease as the sample size increases. 
From Table \ref{vs.sim}, it is also clear that, for fixed dimension $d$, the probabilities decrease as the correlation coefficient increases. There is no clear pattern between inclusion probabilities and the dimension $d$ of the fitted model.
Results for the third mean model choice are available in the supplementary materials. 
They follow the same pattern as the results of Table \ref{vs.sim}, with the probabilities being, as expected, a bit higher.

\begin{table}
\begin{center}
\caption{Simulation study results: the entries of the table are the posterior probabilities, expressed as percentages, that at least one of $x_2, x_3$ is included in the mean model of the first response. 
Rows refer to the dimension of the fitted model $d=2,4,6,10$, columns to the correlation coefficient $\rho=
0.1,0.3,0.5,0.7,0.9$, and the two parts of the table to the two sample sizes $n = 50, 150$.
Results are based on $40$ replicate datasets per sample size by correlation combination.} \label{vs.sim}
\begin{tabular}{lrrrrr|rrrrr}
& \multicolumn{5}{c}{$n=50$} & \multicolumn{5}{c}{$n=150$}\\
\hline
& 0.1 & 0.3 & 0.5 & 0.7 & 0.9 & 0.1 & 0.3 & 0.5 & 0.7 & 0.9\\
\hline
2 & 7.91 & 7.79 & 7.43 & 6.89 & 5.44 & 6.53 & 6.04 & 5.59 & 4.62 & 3.63\\
4 & 8.33 & 8.14 & 7.72 & 7.13 & 5.45 & 6.62 & 6.32 & 5.96 & 5.48 & 4.27\\
6 & 8.48 & 8.38 & 7.90 & 7.36 & 5.93 & 6.53 & 6.27 & 6.13 & 5.73 & 4.62\\
10 & 8.03 & 7.87 & 7.46 & 7.22 & 5.78 & 6.65 & 6.40 & 6.21 & 5.80 & 4.79\\
\hline
\end{tabular}
\end{center}
\end{table}

We conclude this section by reporting run-times of the models. We note that the MCMC sampler has been implemented in the C 
programming language and that the current simulations were run on an Intel Core i7 3.40GHz processor. 
Run-times are reported in Table \ref{time.sim}. These range from $1.64$ sec for a univariate, simple linear regression model
to about $81.5$ min, or $4890$ sec, for a ten-dimensional response model.  
For both sample sizes, and all mean models, increasing the number of responses increases the run-time
in a manner that is consistent with a cubic polynomial. Further, for all response dimensions, 
increasing the sample size increases the run-time linearly. This last point is not obvious from    
Table \ref{time.sim} as there are only two sample sizes, however, this was observed in other simulation studies 
not reported here.   

\begin{table}
\begin{center}
\caption{Simulation study results: the entries of the table are the run-times, measured in seconds, required to obtain 40,000 posterior
samples.  
Rows refer to the sample size $n = 50, 150,$ columns to the dimension of the fitted model $d=1,2,4,6,10$, 
and the three parts of the table to the three mean model specifications.
Results are based on $40$ replicate datasets per sample size by correlation combination.} \label{time.sim}
\begin{tabular}{lrrrrr rrrrr}
& \multicolumn{5}{c}{$\mu_{j} = \beta_{0j} + \beta_{j1} x_{1}$} & \multicolumn{5}{c}{$\mu_{j} = \beta_{0j} + \sum_{k=1}^3 \beta_{jk} x_{k}$} \\		
\hline
& 1 & 2 & 4 & 6 & \multicolumn{1}{r|}{10} & 1 & 2 & 4 & 6 & 10\\
\hline
50 & 1.64 & 10.42 & 25.74 & 59.77 & \multicolumn{1}{r|}{253.38} & 3.07 & 15.29 & 44.60 & 112.34 & 526.81\\
150 & 2.96 & 25.49 & 63.02 & 148.93 & \multicolumn{1}{r|}{639.28} & 5.36 & 36.64 & 106.06 & 270.72 & 1270.04\\
\hline
& \multicolumn{5}{c}{$\mu_{j} = \beta_{0j} + \sum_{k=1}^{10} \beta_{jk} x_{k}$} & \\		
\cline{1-6}
& 1 & 2 & 4 & 6 & 10\\
\cline{1-6}
50 & 9.50 & 37.09 & 147.99 & 466.96 & 2683.43\\
150 & 16.29 & 82.03 & 308.95 & 908.60 & 4890.23\\
\cline{1-6}
\end{tabular}
\end{center}
\end{table}

\section{Applications}\label{applications}

This section describes two applications of the multivariate response model. The first application investigates 
how the human cardiovascular system responds to a particular kind of drug overdose. Due to the complexity of the 
cardiovascular system, a multivariate response measurement has been taken, thus the scientific objectives demand 
flexible regression models within a multivariate framework. The second application shows how the multivariate model can be used 
to semi-parametrically condition on additional information when fitting graphical models. We elaborate on a 
particularly nice example of this type of modelling described in \citet[p.1]{Whittaker}.
The data used in the first application comes from \citet{appliedMSA}. Data for the second application comes from 
\citet{Whittaker} who in turn cites \citet{MKB} as the original source. 

\subsection{Multiple response regression}

The cardiovascular system of $n=17$ patients who had overdosed on amitriptyline (used to treat headaches and depression) 
was measured by taking a blood pressure reading (bp, $y_1$) and also by recording each patients' PRQRS wave - as produced by an 
electrocardiogram. The PRQRS wave was broken down into two parts; the PR part (pr, $y_2$) and the QRS part (qrs, $y_3$). 
Hence, in this example, the number of responses is $p=3$. Covariates include the size of the overdose that was measured in terms of the amount of the drug 
taken (amt), total blood plasma level (tot) and the amount of amitriptyline found inside the plasma 
(ami). The objective of this analysis is to obtain graphical and numerical summaries of the effects of the drug 
overdose, along with a quantification of the uncertainty around those summaries.

To avoid numerical instability as a result of the variables being measured on different scales, we work with centred and scaled 
versions of the responses. In addition, a new covariate defined as $\text{ratio} = \text{ami}/\text{tot}$ is introduced, and the explanatory variables 
are taken to be centred and scaled versions of $\log(\text{amt}), \log(\text{tot}), \log(\text{ratio})$, henceforth simply refer to as amt, tot, and ratio. 
The specific form of the model is
\begin{equation}
\uY_{i} \sim N(\umu(\ux_i,\ubeta^{\ast}), \uSigma(\uR,\usigma^2)), i=1,2,\dots,17,\nonumber
\end{equation}
with the means $\umu(\ux_i,\ubeta^{\ast}) = \left ( \mu(\ux_i,\ubeta_1^{\ast}), \mu(\ux_i,\ubeta_2^{\ast}), \mu(\ux_i,\ubeta_3^{\ast})\right )^{\top}$ given the following shared representation
\begin{equation}
\mu(\ux_i,\ubeta_j^{\ast}) = \beta_{0j} + f_{\mu,j,1}(u_{i1}) + f_{\mu,j,2}(u_{i2}) + f_{\mu,j,3}(u_{i3}), j = 1, 2, 3, \label{share-mean}\nonumber
\end{equation}
where $u_1, u_2$ and $u_3$ denote the three explanatory variables.
The functions $f_{\mu,j,k}$ are represented as  
\begin{equation}
f_{\mu,j,k}(u_{ik}) = \sum_{l=1}^{6} \beta_{jkl} \phi_{\mu kl}(u_{ik}), j = 1, 2, 3, k = 1, 2, 3. \label{basis}\nonumber
\end{equation}
The same number of knots, $5$, or equivalently $6$ basis functions, was chosen for all three semi-parametric terms. 
For each function $f_{\mu,j,k}$, the same $ \pi_{\mu jk} = 0.5$ prior probability for the inclusion of 
$\phi_{\mu kl}(.)$, $j,k = 1, 2, 3, l = 1, \dots,6,$ was used.
These decisions were motivated by not having any reason to want to build in differing levels of functional complexity across the responses, 
nor across the explanatory variables.

Initial plots suggest little to no change in the variances of the response variables, although it is doubtful 
whether the eye or a model would be able to detect this with $n = 17$. For this reason $\uS$ was taken to consist of constant terms
\begin{equation}
\uS = \text{diag} (\sigma_{1}^2, \sigma_{2}^2, \sigma_{3}^2).\nonumber
\end{equation}

The grouped variables prior was placed on $\uR$, with the upper limit $G$ on the number of clusters set to $3$. 
This choice was guided by the the fact that responses pr $(y_2)$ and qrs $(y_3)$ are both measurements of the same biological 
feature (the PRQRS curve) and it would make sense for them to be similarly related to bp $(y_3)$. By choosing $G$ to be equal to the number
of responses, we allow for the possibility that such a grouping is not supported by the data. 

\begin{figure} 
\centering
\begin{tabular}{ccc} 
	\includegraphics[width=40mm]{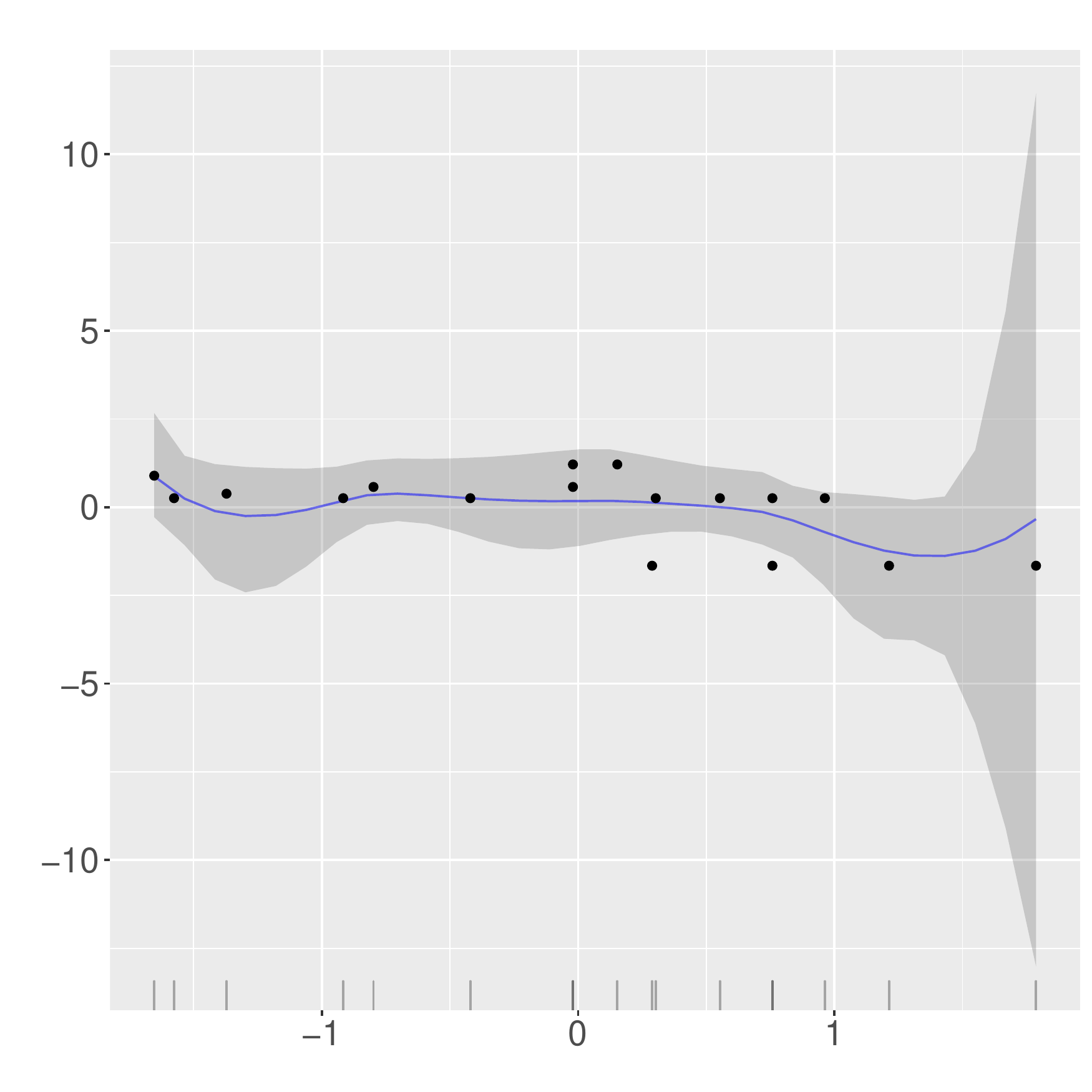} & \includegraphics[width=40mm]{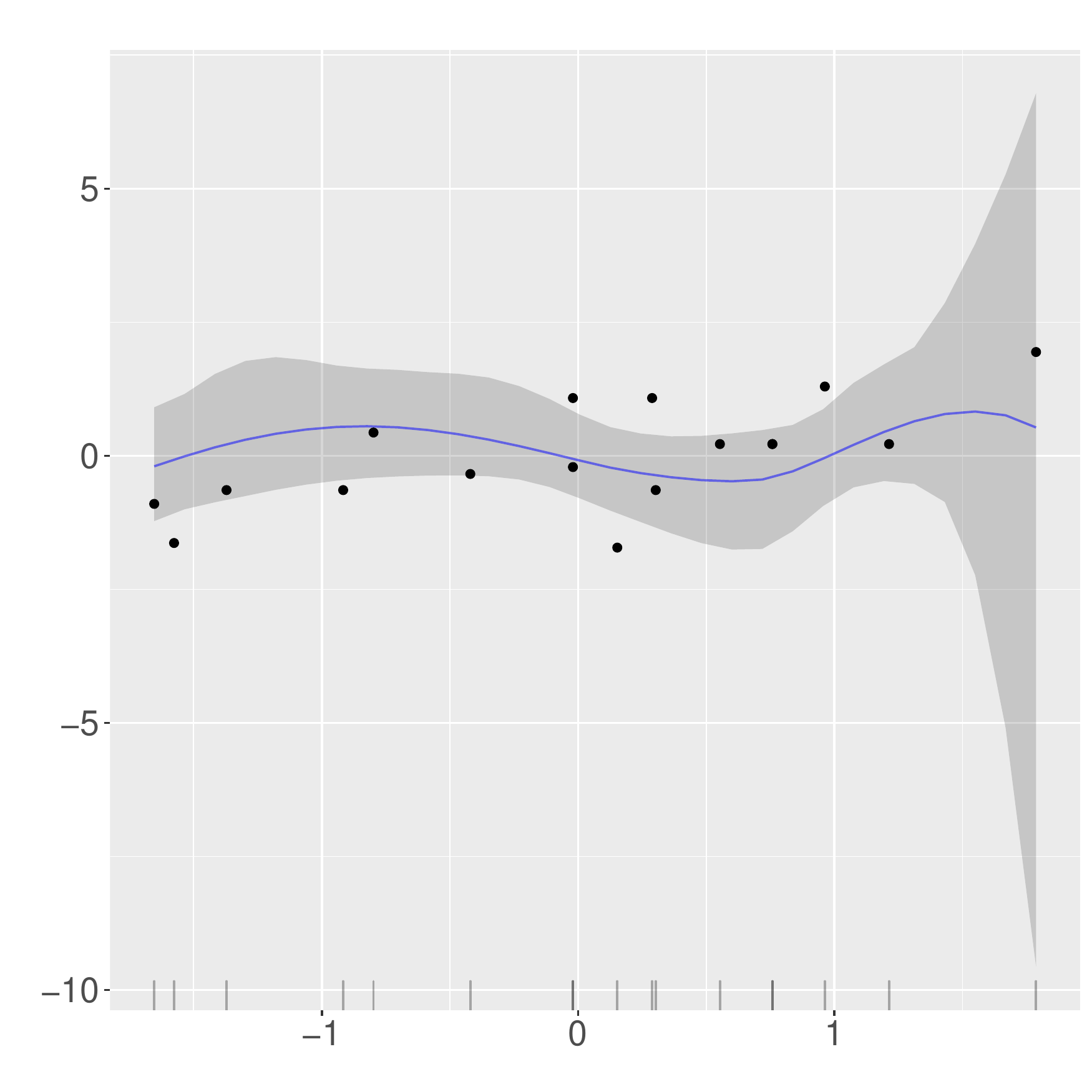} &  \includegraphics[width=40mm]{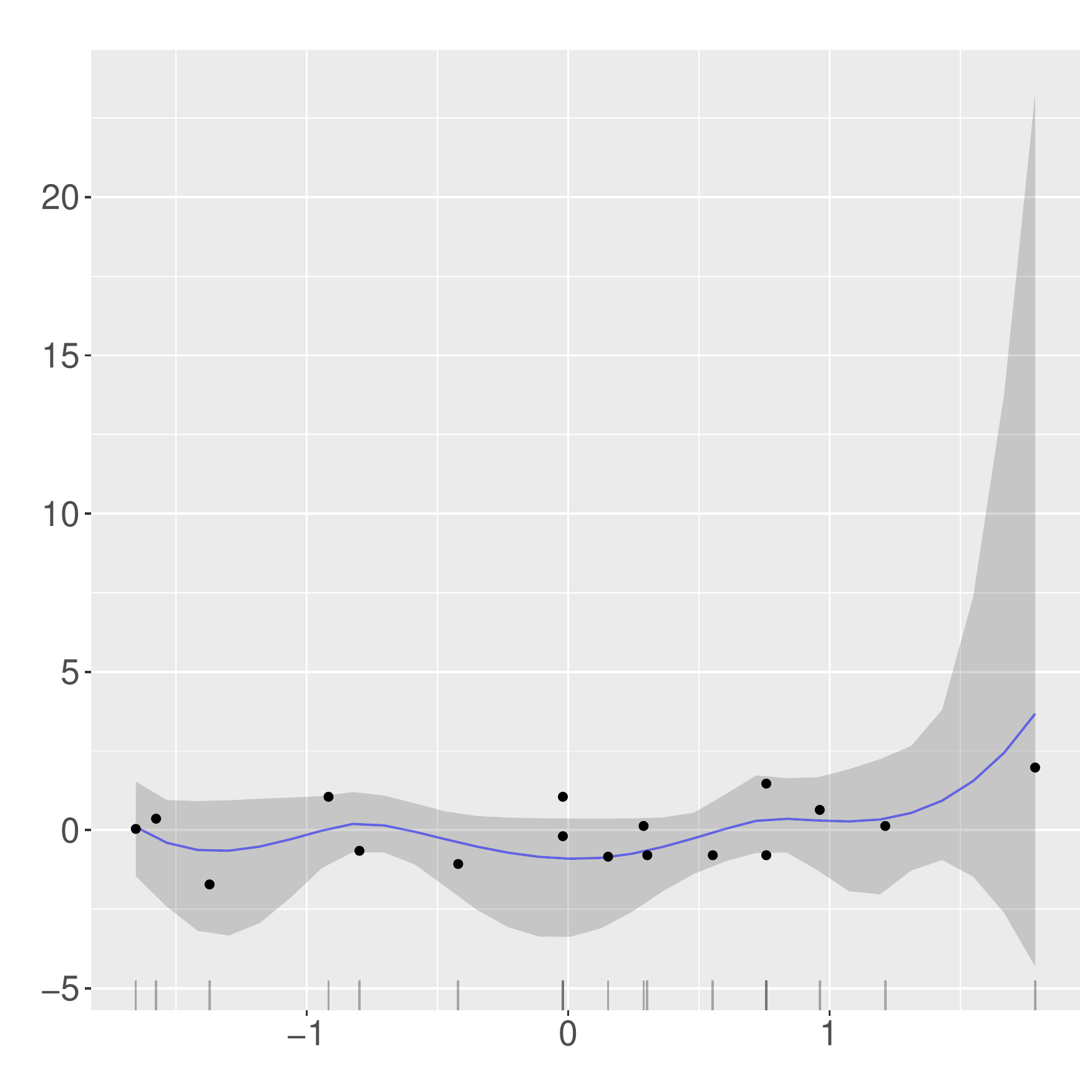}\\
	$f_{\text{bp}}(\text{amt})$ & $f_{\text{pr}}(\text{amt})$ & $f_{\text{qrs}}(\text{amt})$\\
	\includegraphics[width=40mm]{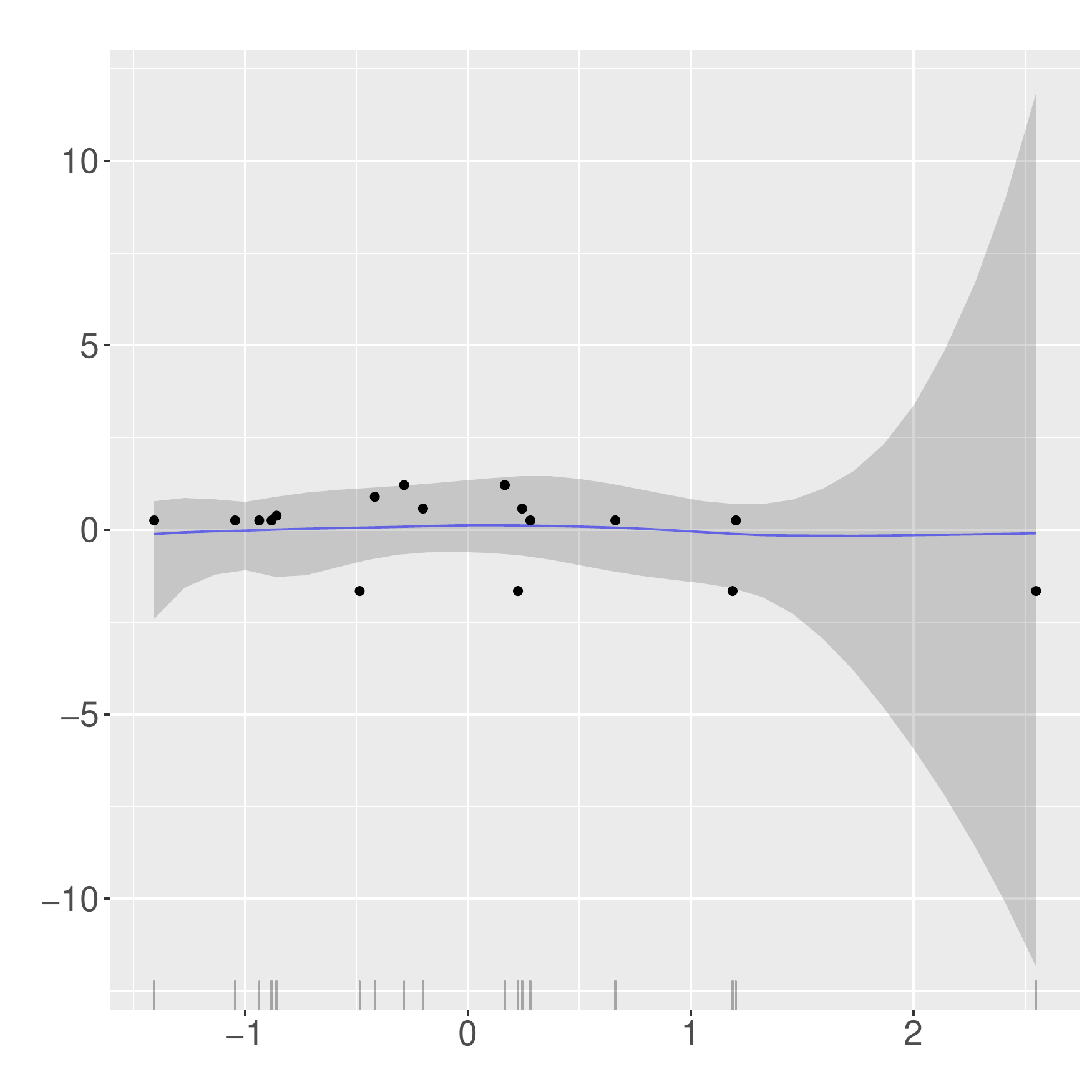} & \includegraphics[width=40mm]{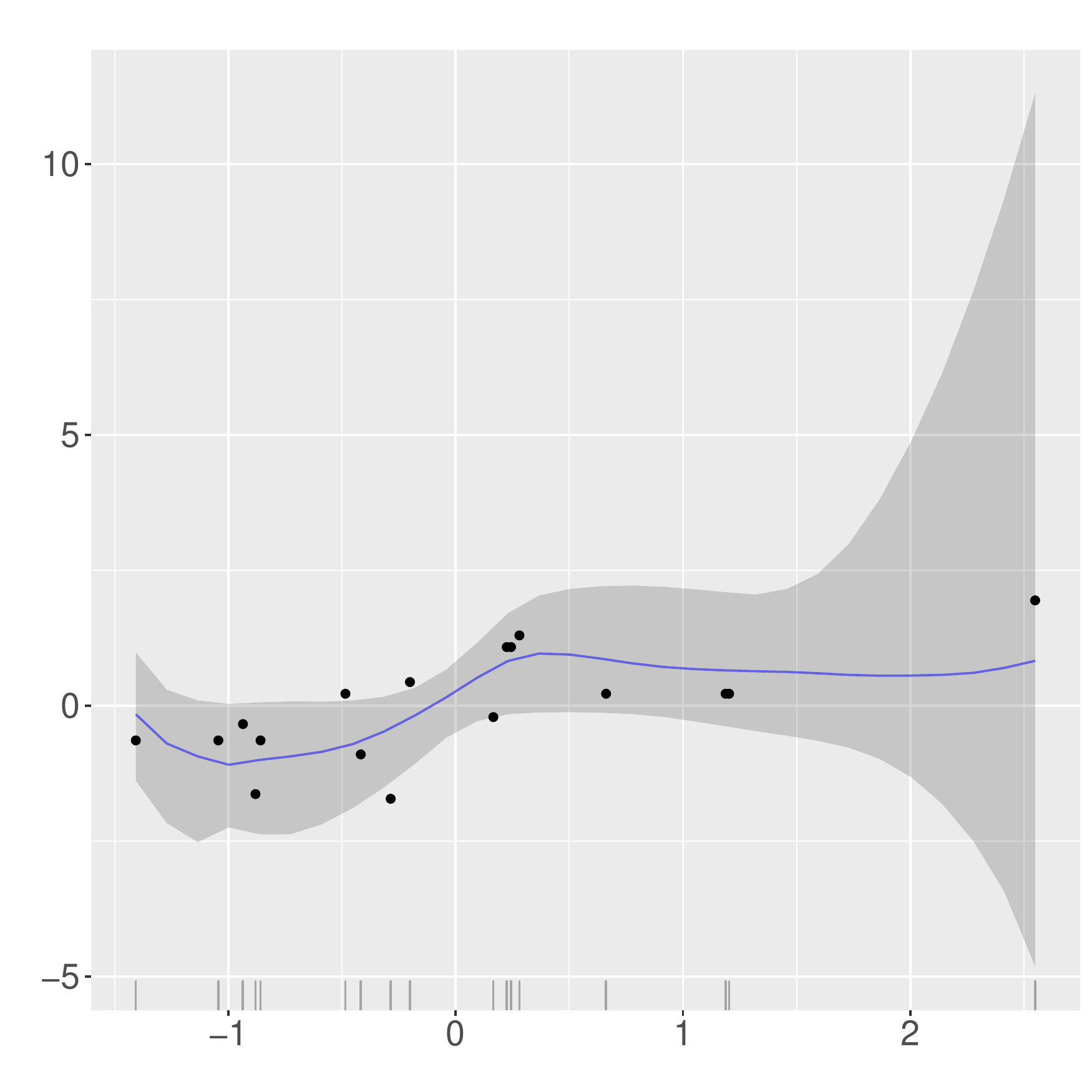} &  \includegraphics[width=40mm]{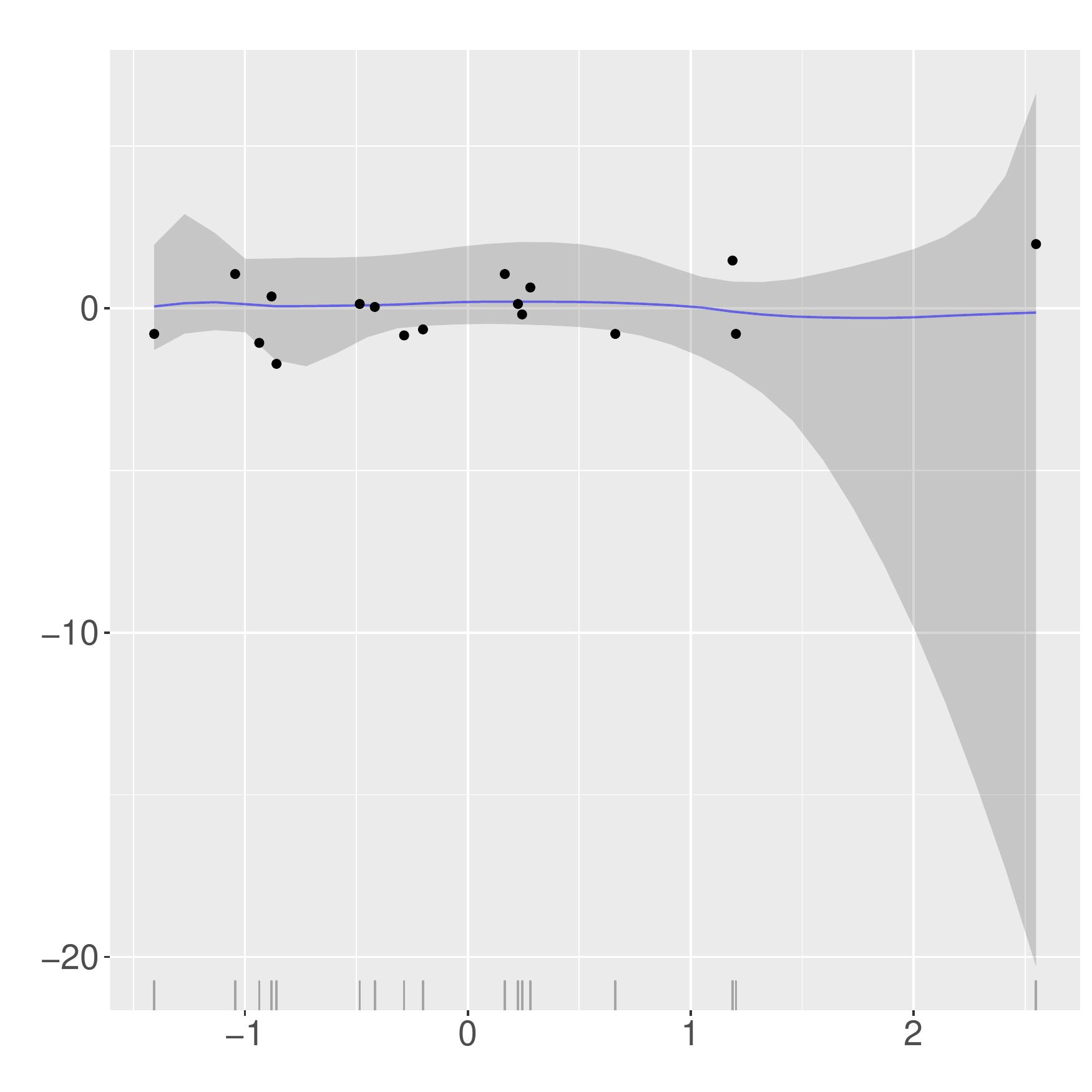}\\
	$f_{\text{bp}}(\text{tot})$ & $f_{\text{pr}}(\text{tot})$ & $f_{\text{qrs}}(\text{tot})$\\
    \includegraphics[width=40mm]{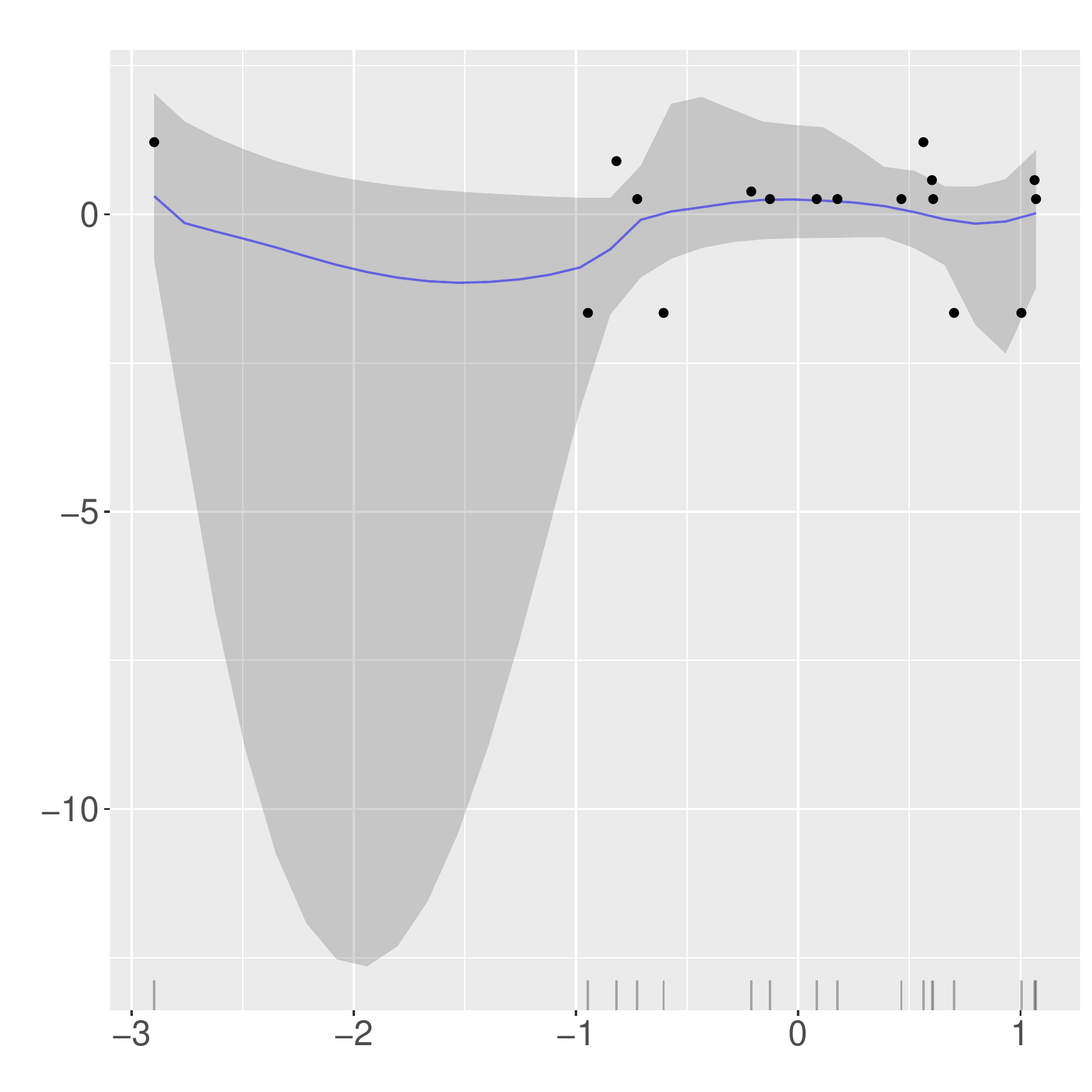} & \includegraphics[width=40mm]{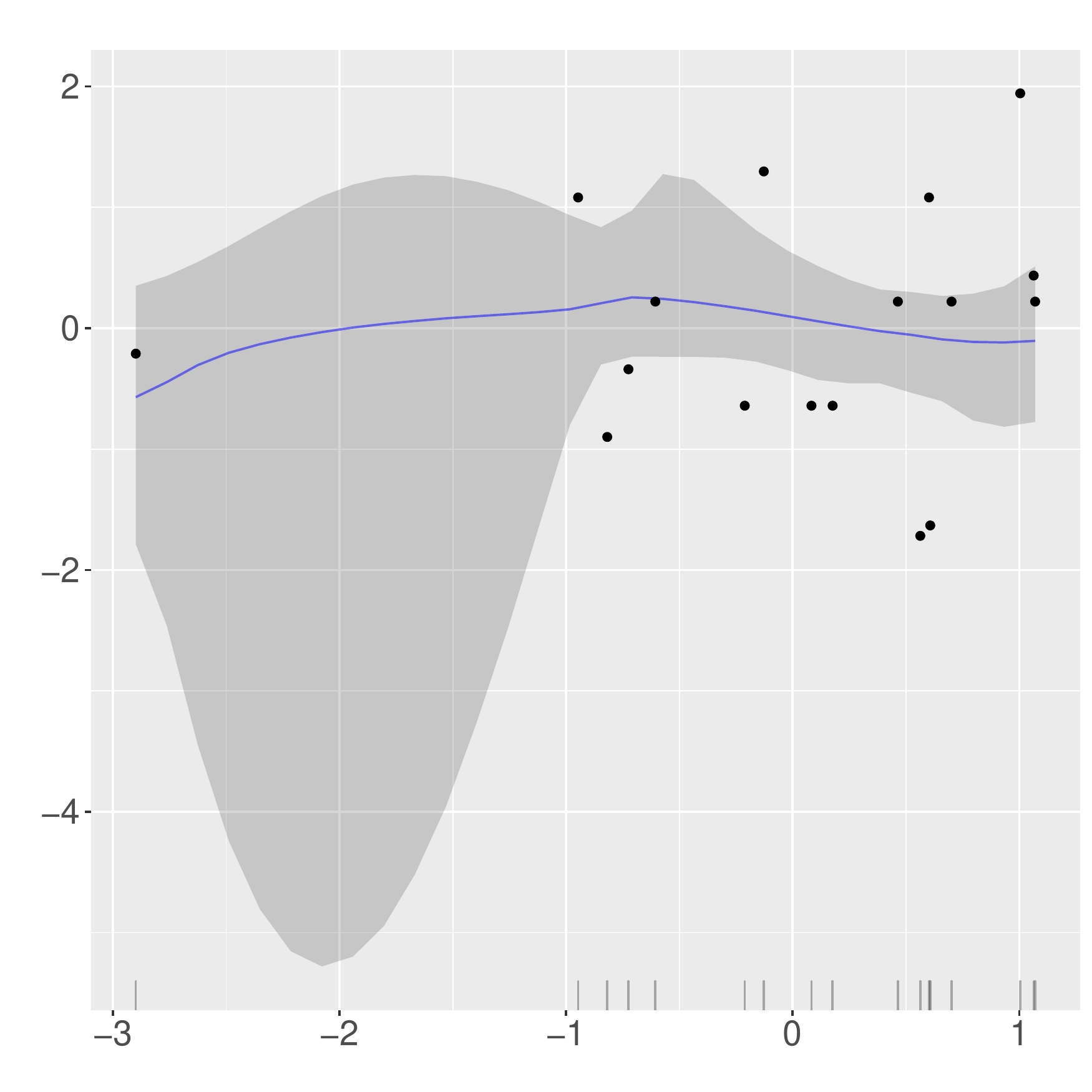} &  \includegraphics[width=40mm]{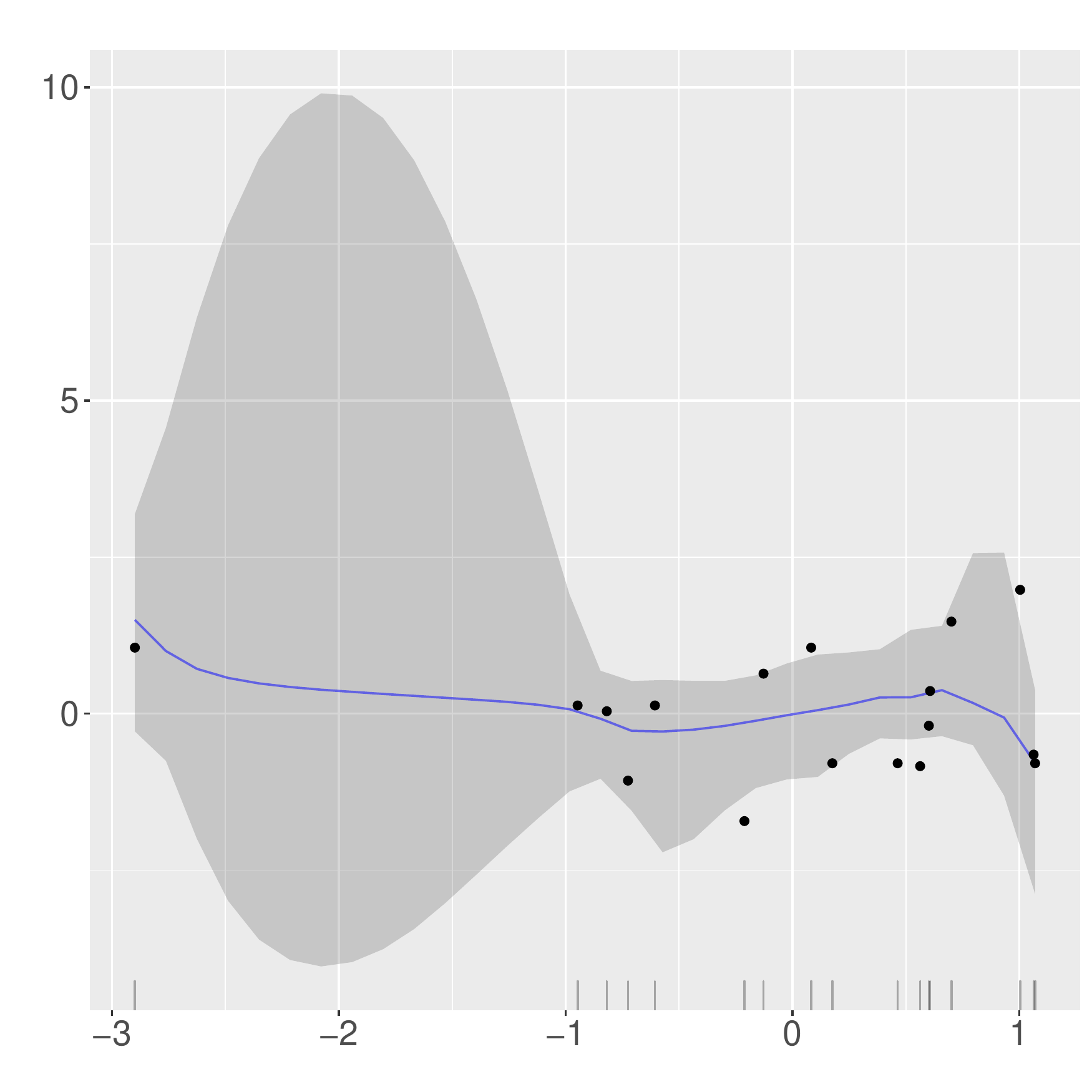}\\
	$f_{\text{bp}}(\text{ratio})$ & $f_{\text{pr}}(\text{ratio})$ & $f_{\text{qrs}}(\text{ratio})$
\end{tabular}
\caption{Results on multiple response regression: posterior means and $90$\% credible intervals over the nonlinear functions that enter the mean models. Rows correspond to the three covariates (amt, tot, ratio) and columns to the three responses (bp, pr, qrs).}\label{postamt}
\end{figure}

The MCMC sampler was run for a total $400,000$ iterations discarding the first $200,000$ as burn in and thereafter retaining every 
second sample. Results are displayed in Figure \ref{postamt}. 
The first row displays the fitted curves for input amt and the three responses, bp, pr, and qrs. 
There is some evidence of nonlinear relationships, with the corresponding $90$\% credible intervals being very wide, reflecting a high 
level of uncertainty due to the high variance in the responses and the small sample size. 
Figure \ref{postamt}, row two, plots the fitted function for covariate tot and the three responses. 
Again, we observe some evidence of nonlinear relationships, with very wide $90$\% credible intervals.
Lastly, the third row plots the fitted functions for covariate ratio. These plots highlight 
the way in which the credible intervals adapt to data sparsity. Where there is less data, the $90$\% credible 
interval is much wider. 

The posterior summaries of the correlations in $\uR$ are given in Table \ref{postcorr}. Displayed are the posterior means, standard deviations, $90\%$ point-wise credible intervals and probabilities of being allocated in the same cluster.
The credible intervals are wide, reflecting the high degree of uncertainty in the values 
of the residual correlations. The posterior over the clustering structure places \text{pr} with \text{qrs} $56 \%$ of the time, and places \text{bp} with \text{pr} and \text{qrs} $50\%$ and $52\%$ of the time, respectively.

\begin{table} 
\centering
\caption{Results on multiple response regression: posterior correlation summary. Rows correspond to the posterior mean, standard deviation, $5\%$ and $95\%$ quantiles, and the probabilities of being allocated in the same cluster. Columns correspond to response variable pairs.}\label{postcorr}	
\begin{tabular}{l|rrr}
	\hline
	& $(\text{pr},\text{qrs})$ & $(\text{pr},\text{bp})$ & $(\text{qrs},\text{bp})$  \\ \hline
	Mean & 0.18 & -0.31 & -0.61  \\
	SD & 0.37 & 0.37 & 0.36  \\ 
	5\% & -0.42 & -0.87 & -0.95 \\
	95\% & 0.77 & 0.26 & 0.17 \\
	Cluster & 0.56 & 0.50 & 0.52 \\
\end{tabular}
\end{table}

\subsection{Graphical Models}

The multivariate normal allows for conditional independence results to be inferred from the structure found in the precision 
(inverse covariance) matrix. In particular, suppose vectors $\uX_a, \uX_b, \uX_c$ are jointly normal. It follows that $\uX_a$ is independent 
of $\uX_b$ given $\uX_c$, if and only if, the (two identical) blocks of precision parameters relating $\uX_a$ with $\uX_b$ are all zero. 
This relation between conditional independence and the precision matrix is proven by considering how the multivariate normal density 
factorises when the precision matrix contains blocks of zeros.

\citet{Whittaker} presents an application of this technique. The data consist of scores on $p=5$ tests given to $n=88$ school children. The tests are mechanics (\text{M}), statistics (\text{S}), vectors (\text{V}), analysis (\text{An}) and algebra (\text{Al}). 
Matrices (a), (b) and (c) in Table \ref{xx1} contain the empirical correlation matrix, scaled negative precision matrix and the suggested independence structure. The independence structure was arrived at by setting to zero all precision terms smaller in absolute value 
than $\alpha = 0.1 $. The same inference would be made for $0.08 < \alpha < 0.23$. 
The interpretation of this structure is that test results on M and V are independent of results on An and S given results on Al. 

Putting aside worries about how to choose a threshold value $\alpha$ in some principled way, we might also wish to explicitly condition on additional information about the school children. If the variables describing this additional information are not normally distributed, then they cannot be added directly into the graphical model. The model presented in this article allows a solution to this problem. We demonstrate this methodology by explicitly conditioning on \text{Al} and repeating the above analysis on the reduced $4 \times 4$ correlation matrix describing the associations between the remaining test results. The analysis described previously suggests we ought to find that there is near zero precision term between the pairs (\text{M}, \text{An}), (\text{M}, \text{S}), (\text{V}, \text{An}) and (\text{V}, \text{S}).    

The model we fit takes the form of 
\begin{equation} 
\uY_{i} \sim N(\umu(\ux_i,\ubeta^{\ast}), \uSigma_i), i=1,2,\dots,88.\nonumber
\end{equation}
Here $\uY_{i} \in \mathbb{R}^4$ is a vector containing the scores (\text{M}, \text{V}, \text{An}, \text{S}) for the $i$th child. The mean vector, $\umu(\ux_i,\ubeta^{\ast})$, is a function of the single explanatory variable $\text{Al}$:
\begin{equation}
\mu(\ux_i,\ubeta_j^{\ast}) = \beta_{0j} + f_{\mu,j}(u_{i}), j = 1, 2, 3, 4,\nonumber
\end{equation}
where $u_{i} = \text{Al}_i$ is the Algebra test score. In this example, there are sufficient data to warrant allowing the variances to vary smoothly with $\text{Al}$. We chose a structure that mirrors the mean model:
\begin{equation}
\log \sigma^2_{ij} = \alpha_{0j} + f_{\sigma,j}(u_{i}), j = 1, 2, 3, 4.\nonumber
\end{equation} 

To complete the specification, a prior needs to be placed over $\uR \in \mathbb{R}^{4 \times 4}$. In light of the objectives of this analysis, 
and motivated by the results obtained previously, we apply the grouped variables prior, with $G=4$, expecting to 
find two groups: $(\text{M}, \text{V})$ and $(\text{An}, \text{S})$.

\begin{figure} 
\centering
\begin{tabular}{cccc} 
		\includegraphics[width=40mm]{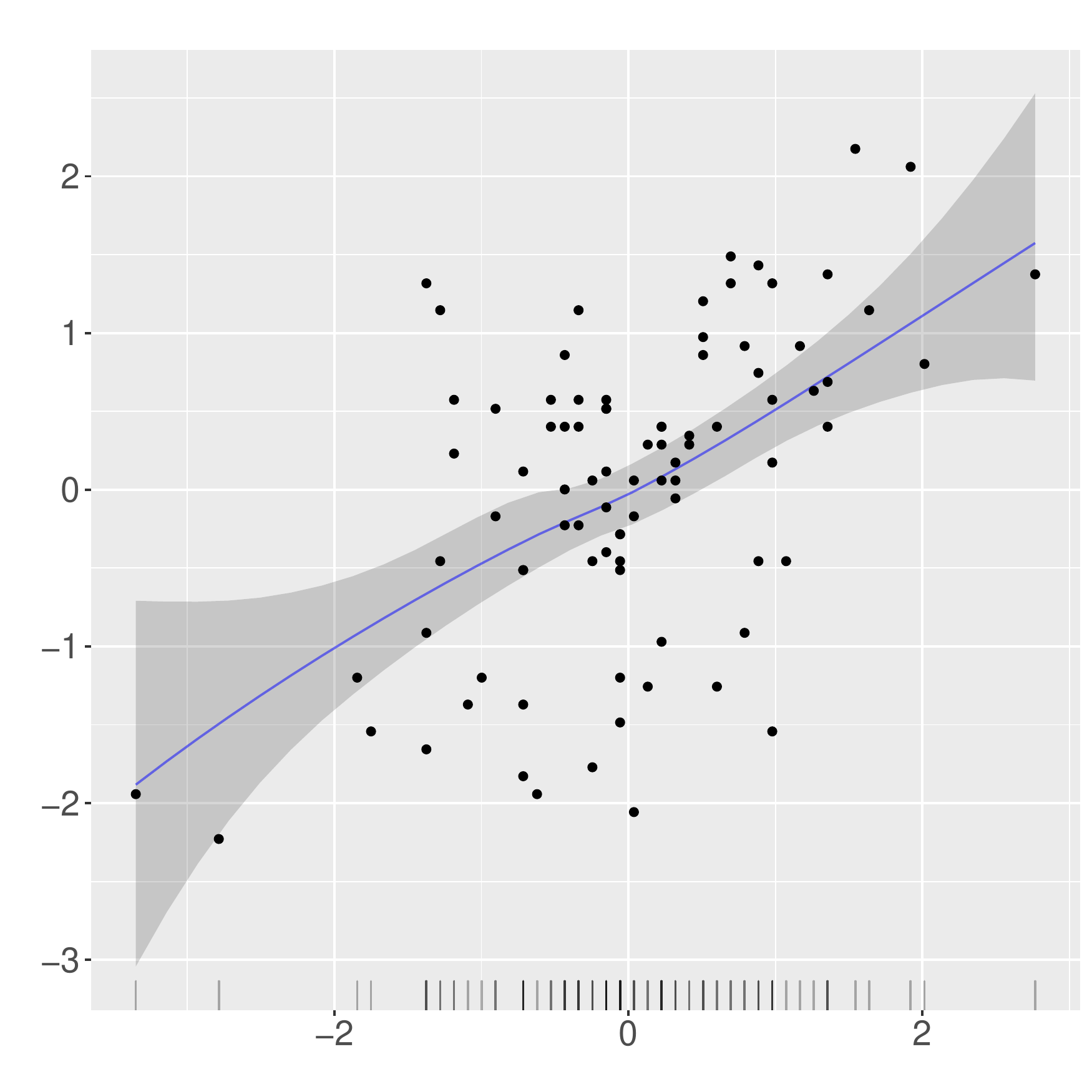} &  \includegraphics[width=40mm]{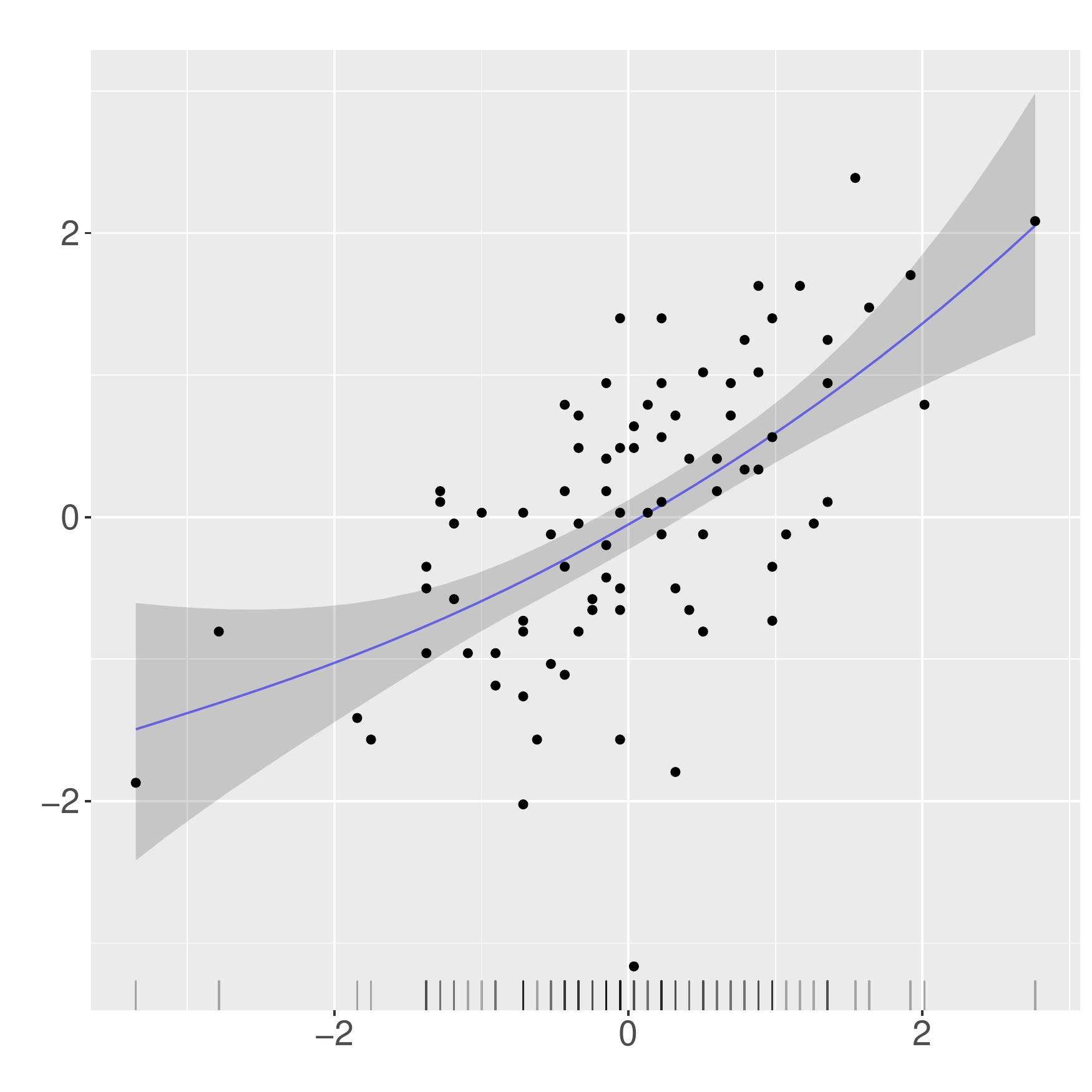} & \includegraphics[width=40mm]{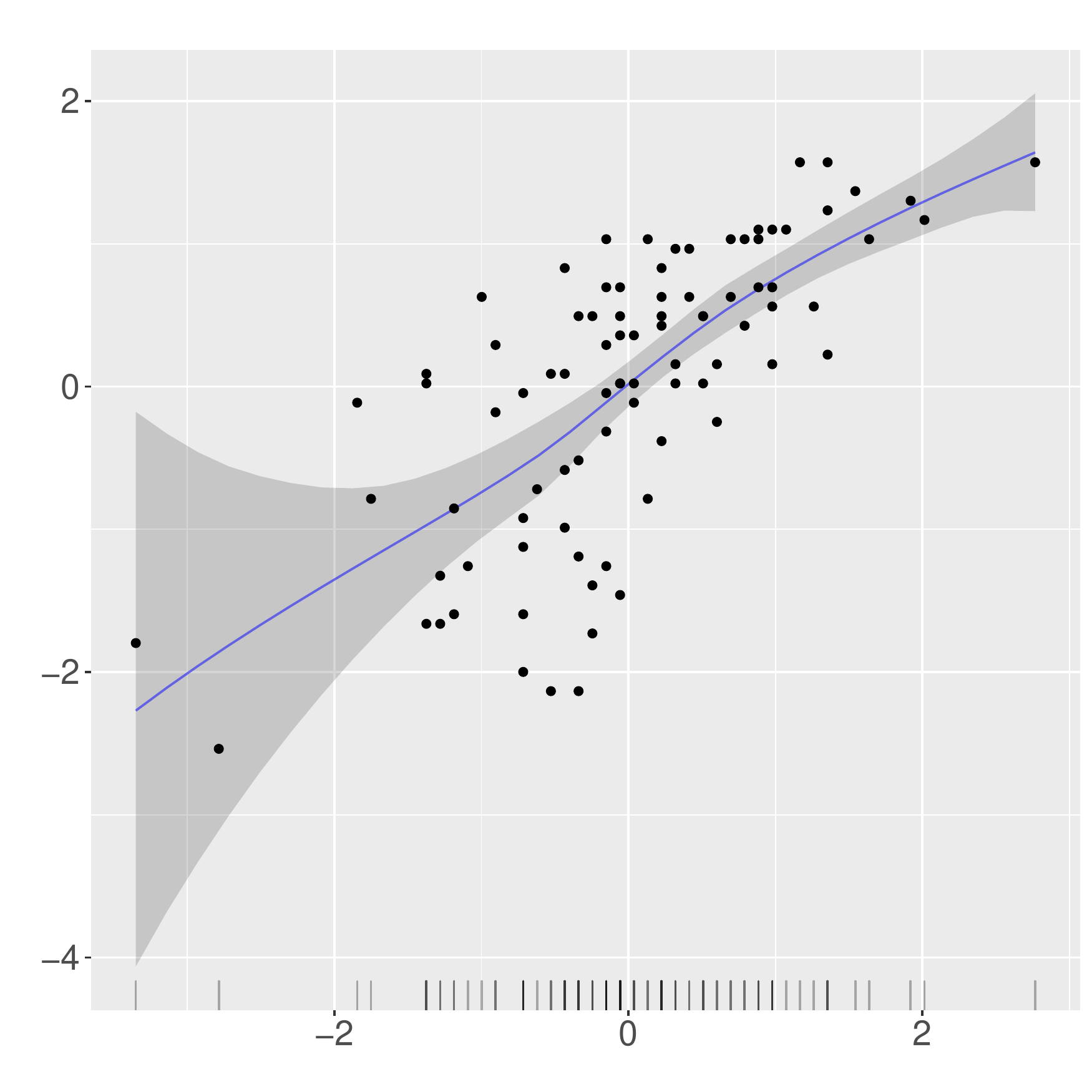} &  \includegraphics[width=40mm]{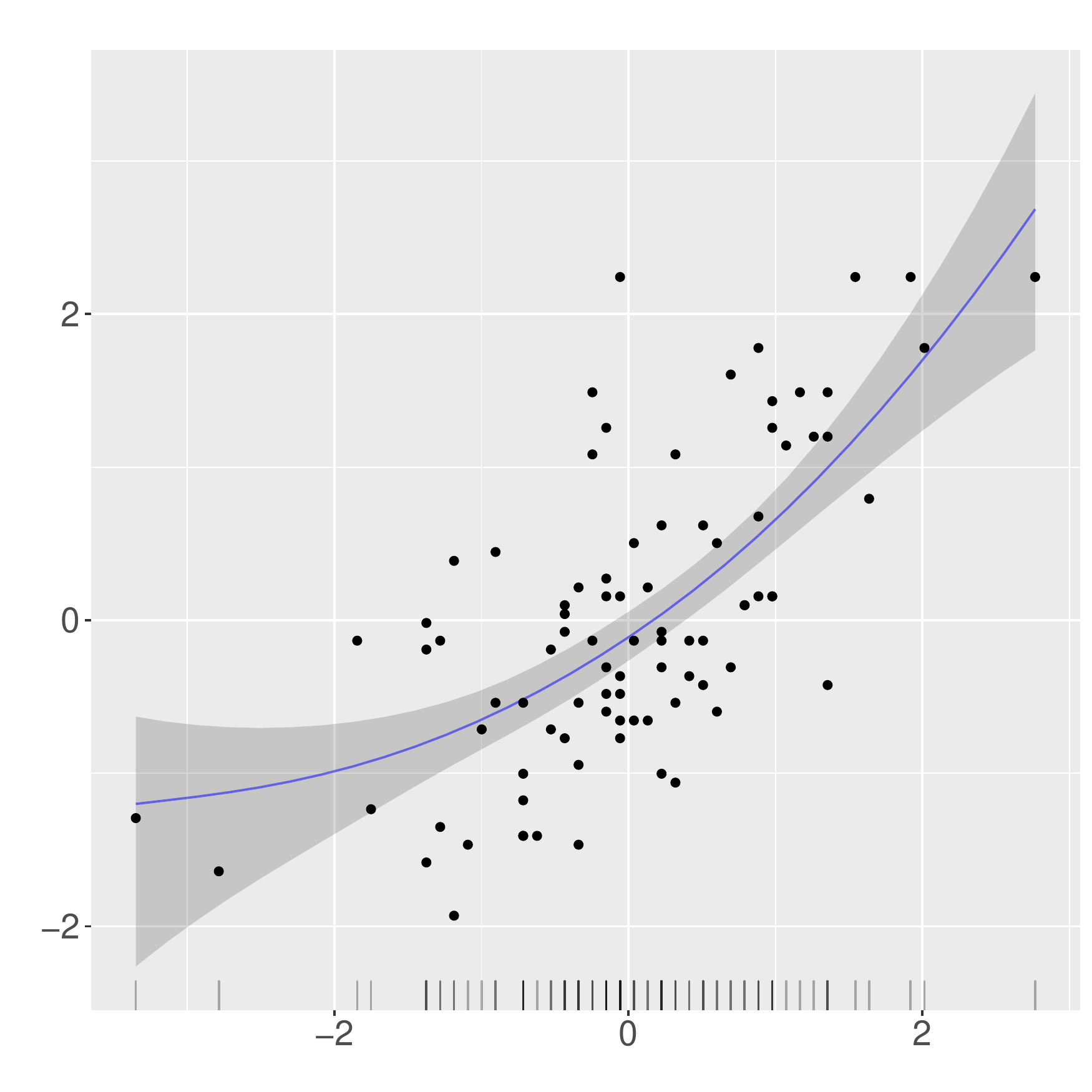} \\
		$\mu_{\text{M}}$ & $\mu_{\text{V}}$ & $\mu_{\text{An}}$ & $\mu_{\text{S}}$\\
		\includegraphics[width=40mm]{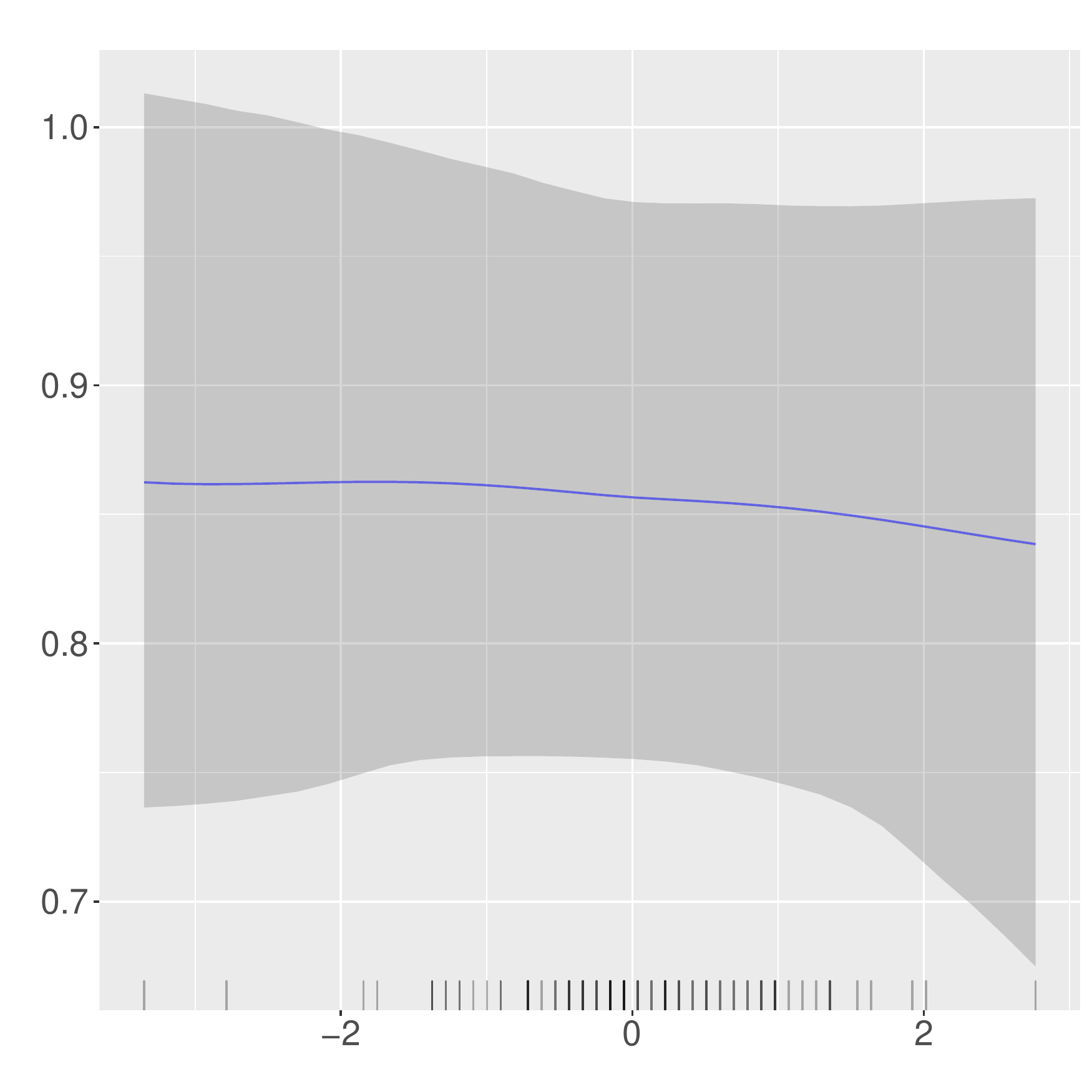} & \includegraphics[width=40mm]{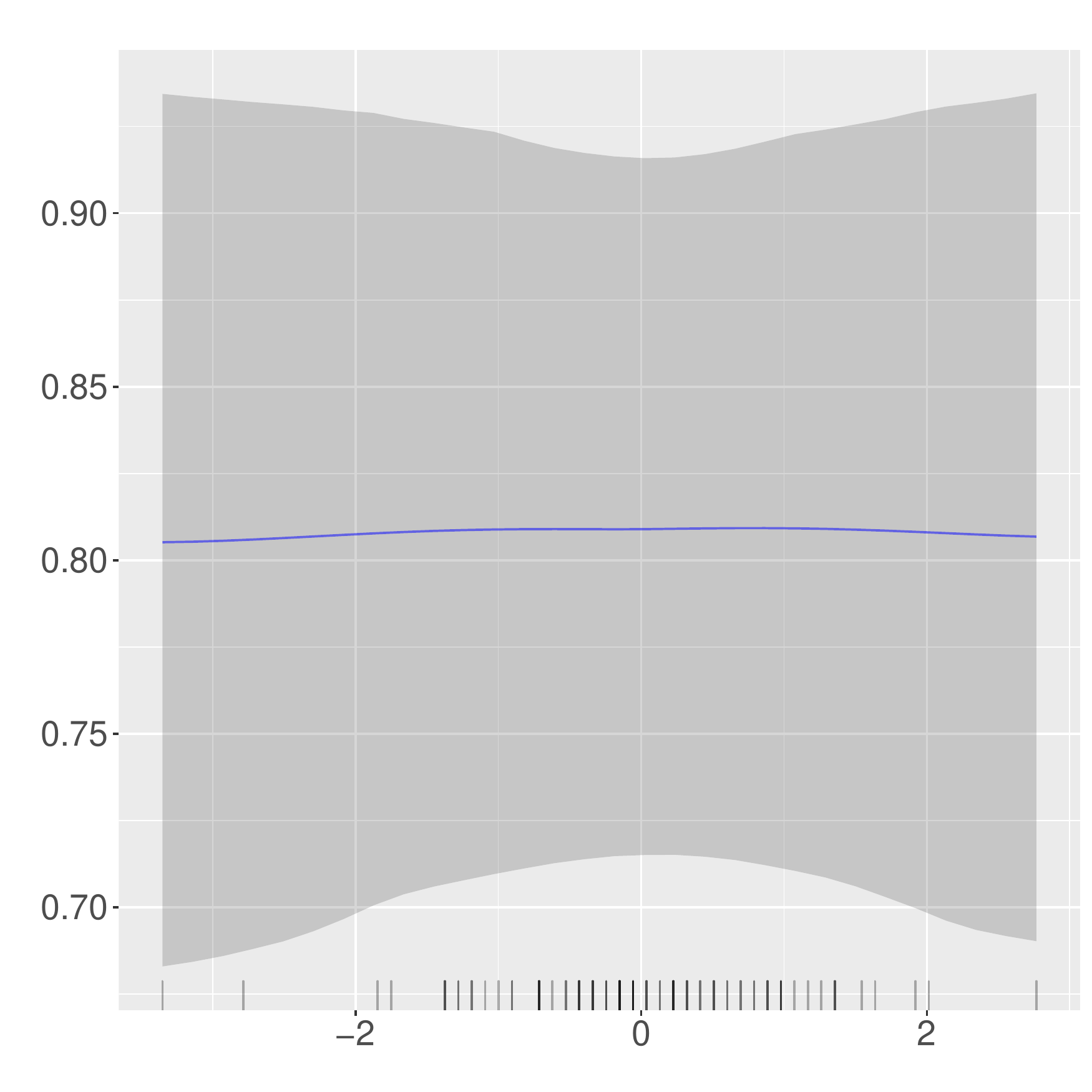} & \includegraphics[width=40mm]{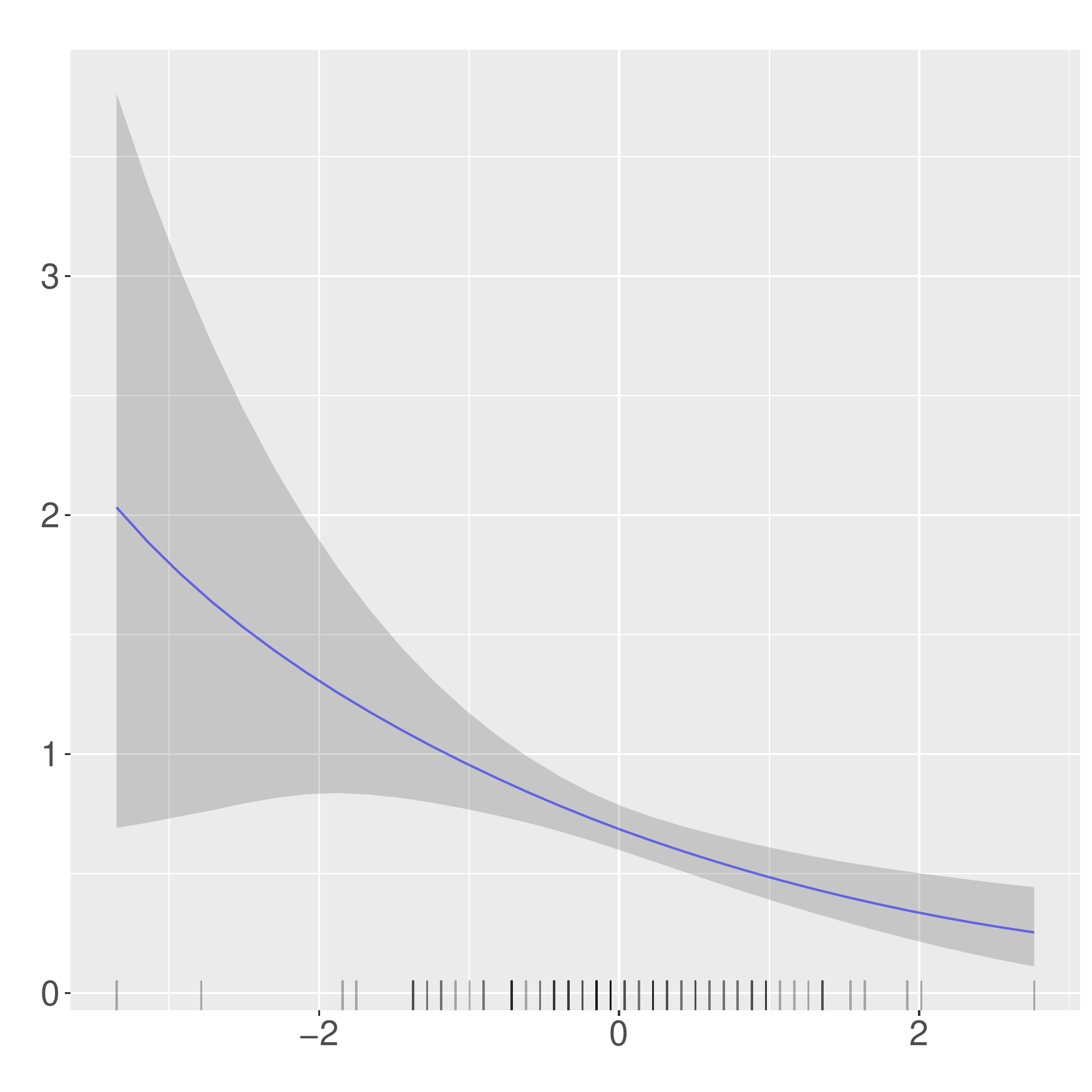} &  \includegraphics[width=40mm]{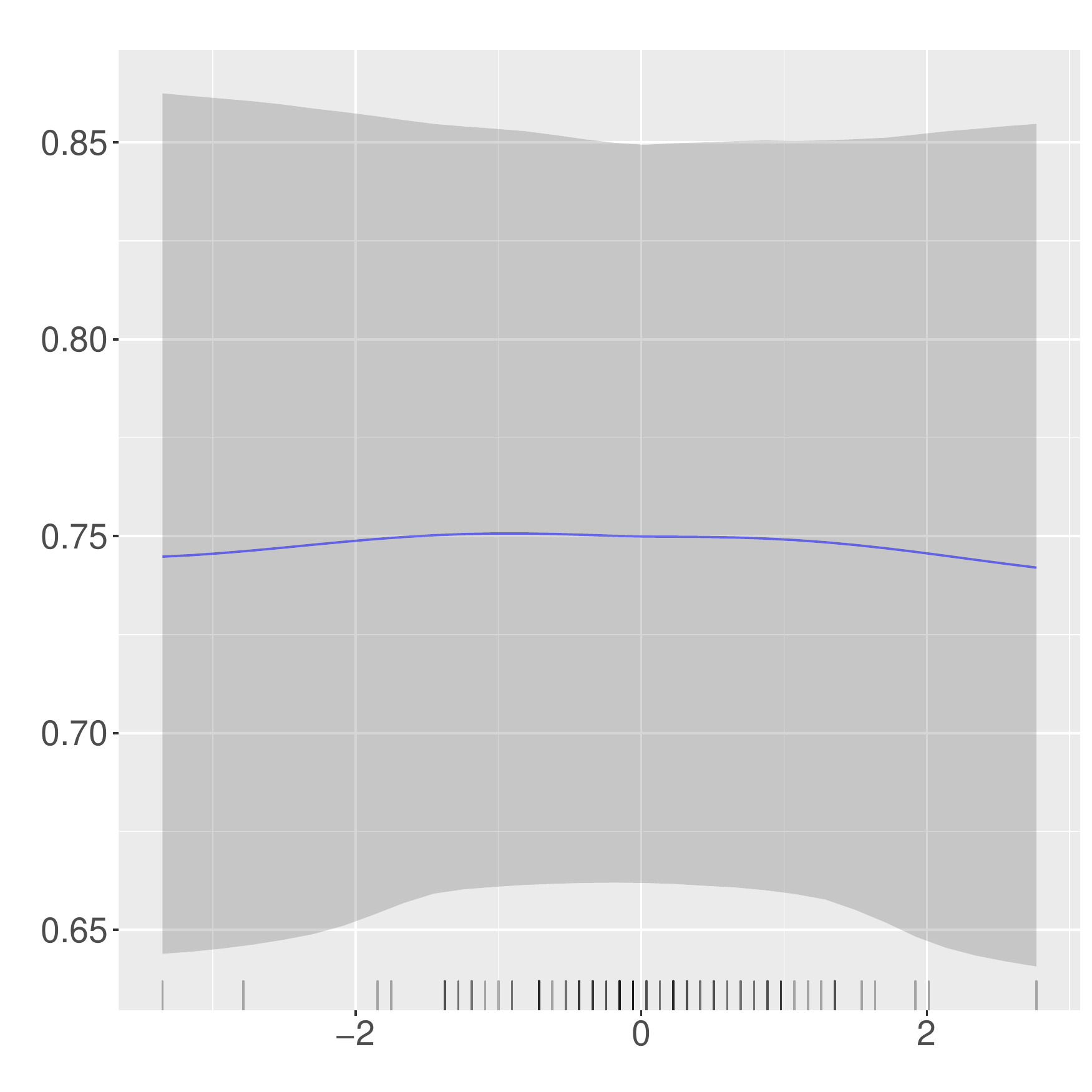} \\ 
		$\sigma_{\text{M}} $ & $\sigma_{\text{V}}$ & $\sigma_{\text{An}}$ & $\sigma_{\text{S}}$
\end{tabular} 
\caption{Results on the graphical modelling application: posterior means and $90\%$ credible intervals of the mean (first row) and standard deviation (second row) functions of the four response variables (M, V, An, S) plotted in the four columns.} \label{pmean}
\end{figure}

The MCMC sampler was run for $400,000$ iterations, discarding $100,000$ as burn in and thereafter retaining every second sample. 
Figure \ref{pmean} presents the estimated functions and $90\%$ credible intervals. There is evidence of non-linear dependency 
of the means on Al. The credible intervals are much tighter in this example, reflecting 
the larger sample size. The credible intervals can also be seen to adapt to the amount of available data. 

The posterior probabilities that the elements of the precision matrix exceed the threshold $\alpha=0.1$ are displayed in Table \ref{xx1},
matrix (d). These are estimated by inverting and scaling every sampled correlation matrix $\uR$, and counting the number of times its elements exceed $\alpha$. The results do conform to a large extent to what was expected. The precision term relating V and M is almost certainly larger
than $\alpha$, with posterior probability essentially one. Likewise, the term relating An with S is greater in magnitude 
than $\alpha$ with probability $0.98$. On the other hand, the terms relating the pairs (An, S) and 
(M, V) all have posterior probabilities of exceeding $\alpha$ far below one. Interestingly, there is still a $0.61$ chance 
that An and V are dependent, even after conditioning on Al, thus displaying the utility of being able to check the assumptions 
behind a graphical model, by explicitly conditioning - in a semiparametric way - on part of the response vector. 

\begin{table} 	
\caption{Results on the graphical modelling application: matrices (a), (b) and (c) are based directly on the analysis given in \citet{Whittaker}. 
	The matrix in (a) is a correlation matrix, and matrices (b) and (c) show the scaled negative precision matrix and the suggested independence structure. 
	Matrix (d) contains the posterior probabilities that the elements of the scaled negative precision matrix are greater  
	than $\alpha = 0.1$ in absolute value, after conditioning on Al.} \label{xx1}	

	\begin{subtable}[h]{0.45\textwidth}
		\centering
		\begin{tabular}{llllll}
			& \text{M} & \text{V} & \text{Al} & \text{An} & \text{S} \\
			\text{M} & 1.00 & & & & \\
			\text{V} & 0.55 & 1.00 & & & \\
			\text{Al}& 0.55 & 0.61 & 1.00 & & \\
			\text{An}& 0.41 & 0.49 & 0.71 & 1.00 & \\
			\text{S} & 0.39 & 0.44 & 0.66 & 0.61 & 1.00 
		\end{tabular}
		\caption{Sample correlation matrix}
	\end{subtable}
	\hfill
	\begin{subtable}[h]{0.45\textwidth}
		\centering
		\begin{tabular}{llllll}
			& \text{M} & \text{V} & \text{Al} & \text{An} & \text{S} \\
			\text{M} &      &      &      &      &      \\
			\text{V} & 0.33 &      &      &      &      \\
			\text{Al}& 0.23 & 0.28 &      &      &      \\
			\text{An}& 0.00 & 0.08 & 0.43 &      &      \\
			\text{S} & 0.03 & 0.02 & 0.36 & 0.25 &      
		\end{tabular}
		\caption{Negative scaled precision matrix}
	\end{subtable}
	\hfill
	\begin{subtable}[h]{0.45\textwidth}
		\centering
		\begin{tabular}{llllll}
			& \text{M} & \text{V} & \text{Al} & \text{An} & \text{S} \\
			\text{M} & 1      &      &      &      &      \\
			\text{V} & 1 & 1     &      &      &      \\
			\text{Al}& 1 & 1   &  1    &      &      \\
			\text{An}& 0 & 0 & 1 &   1   &      \\
			\text{S} & 0 & 0 & 1 & 1 &    1 
		\end{tabular}
		\caption{Independence structure}
	\end{subtable}
	\hfill
	\begin{subtable}[h]{0.45\textwidth}
		\centering
		\begin{tabular}{lllll}
			& \text{M} & \text{V} & \text{An} & \text{S} \\
			\text{M} & 1.00      &         &      &      \\
			\text{V} & 1.00 & 1.00          &      &      \\
			\text{An}& 0.35 & 0.61 &    1.00   &      \\
			\text{S} & 0.22 & 0.21 & 0.98 &    1.00 
		\end{tabular}
		\caption{Independence structure conditioning on Al} 
	\end{subtable}
\end{table}

\section{Discussion}

The article describes a framework for the analysis of multivariate normal responses, with nonparametric models for the means, the variances and the correlation matrix. By utilizing spike-slab priors, the described framework allows covariates that enter the mean and variance functions to automatically drop out of the model. This automatic variable selection can be of great importance when one has to deal with high dimensional datasets.

Our framework builds on the intuitive separation strategy that factorizes the covariance matrix into a diagonal matrix of variances and a correlation matrix. We have described parametric and nonparametric models for the correlation matrix, based on normal and DP mixtures of normals for the (transformed) elements of the correlation matrix. Even though we emphasised DP mixtures in the applications we presented, this certainly is not the only choice. In fact, since the models are intuitive and easy to understand, it is easy for practitioners to incorporate prior knowledge about the correlation structure into the model. 
In a simulation study we illustrated the efficiency gains that one may have when fitting a multivariate models. 
Hence, the method can be useful in practice, since multiple responses naturally arise in many applications. 

\citet{Scheipl12} present a different flavour of spike-slab priors for function selection in univariate structured additive regression models. 
Their model can include varying coefficient terms, smooth interactions between covariates, spatial effects and cluster-specific random effects. 
Allowing for such diverse effects within a multivariate setting is certainly worth pursuing as it  would increase the practical utility of the methods presented here. 

\section{Appendix: MCMC algorithm}\label{mcmc}

At the first step of our sampler, we update the elements of $\ugamma_{jk},j=1,\dots,p,k=1,\dots,K$. This is 
done as suggested by \citet{Chan06}, hence details are omitted, but are available in the supplement. 

At the second step,
pairs $(\udelta_{jk}, \ualpha_{jk}), j=1,\dots,p, k=1,\dots,Q,$ are updated simultaneously. Again, this is done as in   
\citet{Chan06}, who built on the work of \citet{Gamerman1997}, but with the introduction of a free parameter that we select adaptively
\citep{roberts_examples_2009} in order to achieve an acceptance probability 
of $20\% - 25\%$ \citep{Roberts2001c}.
 
The full conditional of $\sigma^2_{j}, j=1,\dots,p,$ is given by
\begin{eqnarray}
f(\sigma^2_{j}|\dots) \propto |\uSigma(\uR,\ualpha,\udelta,\usigma^2)|^{-\frac{1}{2}} \exp(-S/2) \xi(\sigma^2_{j}),\nonumber
\end{eqnarray}
where $\xi(\sigma^2_{j})$ denotes either the IG or half-normal prior. To sample from the above, we follow a random walk algorithm. 
 
The full conditional for parameter $c_{\beta}$ is obtained from the marginal (\ref{marginal}) and the IG$(a_{\beta},b_{\beta})$ prior 
\begin{eqnarray}
	f(c_{\beta}|\dots) \propto (c_{\beta}+1)^{-\frac{N(\gamma)+p}{2}} \exp(-S/2)
	(c_{\beta})^{-a_{\beta}-1} \exp(-b_{\beta}/c_{\beta}).\nonumber
\end{eqnarray}
To sample from the above, we utilize a normal approximation. Let
$\ell(c_{\beta}) = \log\{f(c_{\beta}|\dots)\}$. We utilize a normal proposal density 
$N(\hat{c}_{\beta},-g^2/\ell^{''}(\hat c_{\beta}))$ where $\hat{c}_{\beta}$ is the mode
of $\ell(c_{\beta})$, found using a Newton-Raphson algorithm, 
$\ell^{''}(\hat c_{\beta})$ is the second derivative of $\ell(c_{\beta})$ evaluated at the mode, 
and $g^2$ is a tuning variance parameter that we choose adaptively

Concerning parameter $c_{\alpha j}, j=1,\dots,p$, the full conditional corresponding to the IG$(a_{\alpha j},b_{\alpha j})$ prior is 
another inverse Gamma density, IG$(a_{\alpha j} + N(\delta_j)/2,b_{\alpha j}+\ualpha_{\delta_j j}^{\top}\ualpha_{\delta_j j}/2)$.

Further, using likelihood (\ref{full}) and prior (\ref{gprior}), we find the posterior $\ubeta^{\ast}_{\gamma}$ to be
\begin{eqnarray}
\ubeta_{\gamma}^{\ast} | \dots \sim N(\frac{c_{\beta}}{1+c_{\beta}}(\utX_{\gamma}^{\top} \utX_{\gamma})^{-1}\utX_{\gamma}^{\top} \utY,
\frac{c_{\beta}}{1+c_{\beta}}(\utX_{\gamma}^{\top} \utX_{\gamma})^{-1}).\nonumber
\end{eqnarray}

The next step of the algorithm updates $\uR$. This step has been described in the main body of the article. 

Further, to sample from the full conditional of $\utheta$, write $f(\ur|\utheta,\tau^2) = \nu(\utheta,\tau^2) N(g(\ur);\utheta,\tau^2\uI)$
for the likelihood in (\ref{priorR2}). Further, the prior for $\utheta$ is given in  (\ref{thetaP}), $\utheta \sim N(\mu_R \uone,\sigma^2_R \uI)$. 
Hence, it is easy to show that the posterior is
\begin{eqnarray}\label{ptheta2n}
f(\utheta|\dots) = \nu(\utheta,\tau^2)
N\left(\uA (\tau^{-2} g(\ur) + \sigma^{-2}_R \mu_R \uone),\uA \equiv 
(\tau^{-2} + \sigma_{R}^{-2})^{-1}\uI\right).
\end{eqnarray}
At iteration $u+1$, we sample $\utheta^{(u+1)}$ utilizing as proposal the normal distribution
that appears on the right hand side of (\ref{ptheta2n}).  
The proposed $\utheta^{(u+1)}$ is accepted with probability
\begin{eqnarray}
\min\left\{1,\frac{\nu(\utheta^{(u+1)},\tau^2)}{\nu(\utheta^{(u)},\tau^2)}\right\},\nonumber
\end{eqnarray}
which, for a small value of $\tau^2$ can reasonably be assumed to be unity
\citep{Liechty, Yu201417,Liechty2}. 

We update $\mu_R$ from $\mu_R \sim N((d/\sigma^2_R+1/\varphi^2_r)^{-1} (d/\sigma^2_R)\bar{\theta},(d/\sigma^2_R+1/\varphi^2_r)^{-1})$,
where $\bar{\theta}$ is the mean of the elements of vector $\utheta$. 

Lastly, we update $\sigma^2_R$ utilizing the following full conditional 
\begin{eqnarray}
f(\sigma^2_R|\dots) \propto (\sigma_R^2)^{-\frac{d}{2}} \exp\{-\sum_{i=1}^d(\theta_i-\mu_R)^2 /(2\sigma_R^2)\} \exp\{-\sigma_R^2/(2\phi^2_{R})\} I[\sigma_R>0].\nonumber
\end{eqnarray}
Proposed values are obtained from $(\sigma^2_{R})^{(p)} \sim N((\sigma^2_{R})^{(c)},f_1^2)$ where $(\sigma^2_{R})^{(c)}$ denotes the current value and $f_1^2$ denotes a tuning parameter.

\section{Supplementary Materials}

\begin{description}
	
	\item[Supplement]: Additional tables with simulation results and a detailed MCMC sampler. (.pdf file)
	
	\item[R package BNSP]: An R package that implements the MCMC algorithm and various functions for 
	processing the posterior samples. The package is also available on \href{https://CRAN.R-project.org/package=BNSP}{CRAN}. (.tar.gz)
	
	\item[Examples]: Folder containing R scripts for replicating simulations and data examples. 
	
\end{description}

\bibliographystyle{apalike}
\bibliography{all.bib}

\end{document}


\maketitle

\section{Additional simulation results}

Tables \ref{bias2.sim} and \ref{var2.sim} present results on relative bias and variance for the second choice of the mean model, $\mu_{j} = \beta_{0j} + 
\sum_{k=1}^3 \beta_{jk} x_{k}$. We can observe that the patterns of relative bias and variance are the same as those seen in the main 
body of the paper for the first mean model choice, however, the gains for the second model are generally more pronounced.

Tables \ref{bias3.sim} and \ref{var3.sim} present relative bias and variance results for the third choice of the mean model, 
$\mu_{j} = \beta_{0j} + \sum_{k=1}^{10} \beta_{jk} x_{k}$. The relative bias decreases as $\rho$ increases, for all $n$ and $d$.
This decrease, for $n=50$, is slower than that observed for the first and second mean models, while for $n=150$, 
it is faster than that observed for the first and second mean models. The relative variance follows the same pattern as that observed for 
the other two mean models: it decreases as $\rho$ increases, for all $n$ and $d$. This decrease is faster than
that observed for the first and second mean models. 

Table \ref{comp.sim2} presents the comparison with the MRCE method for the third mean model. 
The proposed method has bias that is usually less than half of the bias of the MRCE approach. 

Table \ref{vs.sim2} presents results on the variable selection performance of the proposed model for the third mean model. 
For this case, when fitting one-dimensional models, the irrelevant regressors where included $9.81\%$ of the time when $n=50$, 
and $6.87\%$ of the time when $n=150$. From this observation and from Table \ref{vs.sim2} we can see that probabilities of inclusion decrease as the sample size increases
and that, for fixed dimension $d$, the probabilities decrease as the correlation coefficient increases. 

\begin{table}
	\begin{center}
		\caption{Simulation study results: the entries of the table are the relative biases $B(d)/B(1), d=2,4,6,10$, 
			expressed as percentages. 
			Rows refer to the sample size $n = 50, 150,$ and columns to the correlation between the responses, $\rho=
			0.1,0.3,0.5,0.7,0.9$. Results are based on the second mean model, and $40$ replicate datasets per sample size by correlation combination.} \label{bias2.sim}
		\begin{tabular}{l|ccccc|l|ccccc}
			$d=2$  &  0.1   &    0.3  &   0.5  &   0.7  &   0.9 & $d=4$  &  0.1   &    0.3  &   0.5  &   0.7  &   0.9\\
			\hline
			50  & 91.85 & 83.54 & 74.45 & 62.65 & 48.83 & 50  & 92.17 & 78.90 & 65.24 & 54.41 & 45.88\\
			150 & 98.26 & 95.02 & 91.95 & 88.25 & 78.35 & 150 & 98.79 & 99.44 & 95.10 & 88.32 & 78.06\\
			\hline
			$d=6$  &  0.1   &    0.3  &   0.5  &   0.7  &   0.9 & $d=10$  &  0.1   &    0.3  &   0.5  &   0.7  &   0.9\\
			\hline
			50  & 97.90 & 77.97 & 63.69 & 53.56 & 45.45 & 50  & 95.21 & 78.98 & 61.84 & 52.34 & 45.04 \\
			150 & 99.72 & 99.84 & 94.98 & 87.50 & 77.10 & 150 & 97.48 & 95.52 & 91.48 & 84.96 & 76.16\\
		\end{tabular}
	\end{center}
\end{table}

\begin{table}
	\begin{center}
		\caption{Simulation study results: the entries of the table are the relative variances $V(d)/V(1), d=2,4,6,10$, 
			expressed as percentages. 
			Rows refer to the sample size $n = 50,150,$ and columns to the correlation between the responses, $\rho=
			0.1,0.3,0.5,0.7,0.9$. Results are based on the second mean model, and $40$ replicate datasets per sample size by correlation combination.} \label{var2.sim}
		\begin{tabular}{l|ccccc|l|ccccc}
			$d=2$  &  0.1   &    0.3  &   0.5  &   0.7  &   0.9 & $d=4$  &  0.1   &    0.3  &   0.5  &   0.7  &   0.9\\
			\hline
			50  & 100.07 & 96.25 & 88.08 & 78.23 & 61.34 & 50  & 100.72 & 92.36 & 79.99 & 67.29 & 52.25\\
			150 & 99.60 &  93.80 & 85.71 & 73.44 & 55.65 & 150 &  99.17 & 90.33 & 79.34 & 67.28 & 52.27\\
			\hline
			$d=6$  &  0.1   &    0.3  &   0.5  &   0.7  &   0.9 & $d=10$  &  0.1   &    0.3  &   0.5  &   0.7  &   0.9\\
			\hline
			50  & 99.19 &  89.42 & 76.23 & 64.52 & 53.94 & 50 &  94.92 & 84.59 & 74.53 & 65.06 & 52.55\\
			150 & 98.65 & 88.43 & 77.53 & 65.44 & 51.61 & 150 & 97.26 & 86.97 & 76.49 & 63.93 & 51.49\\
		\end{tabular}
	\end{center}
\end{table}

\begin{table}
	\begin{center}
		\caption{Simulation study results: the entries of the table are the relative biases $B(d)/B(1), d=2,4,6,10$, 
			expressed as percentages. 
			Rows refer to the sample size $n = 50, 150,$ and columns to the correlation between the responses, $\rho=
			0.1,0.3,0.5,0.7,0.9$. Results are based on the third mean model, and $40$ replicate datasets per sample size by correlation combination.} \label{bias3.sim}
		\begin{tabular}{l|rrrrr|l|rrrrr}
			\hline
			$d=2$  &  0.1   &    0.3  &   0.5  &   0.7  &   0.9 & $d=4$  &  0.1   &    0.3  &   0.5  &   0.7  &   0.9\\
			\hline
			50  & 97.49 & 97.11 & 95.34 & 84.16 & 63.14 & 50 & 99.82 & 91.03 & 84.68 & 74.34 & 58.50\\
			150 & 100.49 & 91.53 & 80.04 & 58.41 & 34.69 & 150 & 102.94 & 93.67 & 81.13 & 57.69 & 33.55\\
			\hline
			$d=6$  &  0.1   &    0.3  &   0.5  &   0.7  &   0.9 & $d=10$  &  0.1   &    0.3  &   0.5  &   0.7  &   0.9\\
			\hline
			50  & 102.88 & 94.32 & 83.82 & 74.63 & 58.47  & 50 & 107.27 & 99.47 & 88.81 & 77.23 & 59.93\\
			150 & 98.80 & 88.18 & 75.77 & 55.32 & 33.08 & 150 & 98.79 & 83.87 & 70.26 & 52.83 & 32.19\\
			\hline 
		\end{tabular}
	\end{center}
\end{table}

\begin{table}
	\begin{center}
		\caption{Simulation study results: the entries of the table are the relative variances $V(d)/V(1), d=2,4,6,10$, 
			expressed as percentages. 
			Rows refer to the sample size $n = 50,150,$ and columns to the correlation between the responses, $\rho=
			0.1,0.3,0.5,0.7,0.9$. Results are based on the third mean model, and $40$ replicate datasets per sample size by correlation combination.} \label{var3.sim}
		\begin{tabular}{l|ccccc|l|ccccc}
			$d=2$  &  0.1   &    0.3  &   0.5  &   0.7  &   0.9 & $d=4$  &  0.1   &    0.3  &   0.5  &   0.7  &   0.9\\
			\hline
			50 & 99.82 & 96.17 & 86.11 & 70.01 & 46.72 & 50 & 100.52 & 93.58 & 79.00 & 60.27 & 38.11\\
			150 & 99.71 & 94.82 & 83.19 & 67.40 & 47.07 & 150 & 100.35 & 88.92 & 73.07 & 54.95 & 36.49\\
			\hline
			$d=6$  &  0.1   &    0.3  &   0.5  &   0.7  &   0.9 & $d=10$  &  0.1   &    0.3  &   0.5  &   0.7  &   0.9\\
			\hline
			50 & 99.90 & 89.40 & 74.20 & 56.48 & 37.92 & 50 & 97.65 & 85.21 & 71.05 & 54.65 & 36.13\\
			150 & 98.06 & 84.78 & 68.21 & 51.49 & 35.61 & 150 & 97.81 & 84.20 & 67.20 & 51.46 & 35.15\\
		\end{tabular}
	\end{center}
\end{table}

\begin{table}
	\begin{center}
		\caption{Simulation study results: the entries of the table are the relative biases $B(d)/B_M(d), d=2,4,6,10$, 
			expressed as percentages. 
			Rows refer to the sample size $n = 50, 150,$ and columns to the correlation between the responses, $\rho=
			0.1,0.3,0.5,0.7,0.9$. Results are based on the third mean model, and $40$ replicate datasets per sample size by correlation combination.} \label{comp.sim2}
		\begin{tabular}{l|rrrrr|l|rrrrr}
			\hline
			$d=2$  &  0.1   &    0.3  &   0.5  &   0.7  &   0.9 & $d=4$  &  0.1   &    0.3  &   0.5  &   0.7  &   0.9\\
			\hline
			50  & 42.57 & 39.61 & 45.52 & 50.37 & 49.86 & 50  & 29.88 & 33.41 & 42.21 & 51.35 & 61.39\\
			150 & 55.00 & 53.50 & 48.59 & 34.37 & 27.48 & 150 & 56.49 & 56.42 & 36.55 & 38.20 & 35.64 \\
			\hline
			$d=6$  &  0.1   &    0.3  &   0.5  &   0.7  &   0.9 & $d=10$  &  0.1   &    0.3  &   0.5  &   0.7  &   0.9\\
			\hline
			50  & 27.44 & 33.36 & 43.75 & 56.66 & 58.28 & 50 & 30.59 & 35.55 & 43.79 & 48.56 & 58.47\\
			150 & 44.92 & 42.89 & 35.81 & 34.62 & 30.78 & 150 & 36.44 & 22.77 & 26.53 & 34.61 & 31.07\\
			\hline 
		\end{tabular}
	\end{center}
\end{table}

\begin{table}
	\begin{center}
		\caption{Simulation study results: the entries of the table are the posterior probabilities, expressed as percentages, that at least one of $x_2, \dots, x_{10}$ is included in the mean model of the first response. 
			Rows refer to the dimension of the fitted model $d=2,4,6,10$, columns to the correlation coefficient $\rho=
			0.1,0.3,0.5,0.7,0.9$, and the two parts of the table to the two sample sizes $n = 50, 150$.
			Results are based on $40$ replicate datasets per sample size by correlation combination.} \label{vs.sim2}
		\begin{tabular}{lrrrrr|rrrrr}
			& \multicolumn{5}{c}{$n=50$} & \multicolumn{5}{c}{$n=150$}\\
			\hline
			& 0.1 & 0.3 & 0.5 & 0.7 & 0.9 & 0.1 & 0.3 & 0.5 & 0.7 & 0.9\\
			\hline
			2 & 10.13 & 10.08 & 9.80 & 9.18 & 7.09 & 6.60 & 6.58 & 6.33 & 5.82 & 4.42\\
			4 & 10.47 & 10.46 & 10.14 & 9.65 & 7.77 & 6.79 & 6.60 & 6.32 & 5.85 & 4.46\\
			6 & 10.86 & 10.68 & 10.30 & 9.90 & 8.03 & 6.67 & 6.43 & 6.17 & 5.77 & 4.64\\
			10 & 11.03 & 10.90 & 10.72 & 10.35 & 9.06 & 6.81 & 6.71 & 6.47 & 6.23 & 5.13\\
			\hline
		\end{tabular}
	\end{center}
\end{table}

\newpage

\section{MCMC algorithm}

Here we provide all details of the MCMC sampler of the three correlation models. 

\subsection{MCMC algorithm for the common correlations model}

Starting from the common correlations model, the algorithm proceeds as follows:
\begin{enumerate}
	
	
\item 
	
As suggested by \citet{Chan06}, the elements of $\ugamma_{jk},j=1,\dots,p,k=1,\dots,K,$ 
are updated in random order and in blocks of random size. Let $\ugamma_{Bjk}$ be a block
of elements of $\ugamma_{jk}$.
The proposed value for $\ugamma_{Bjk}$ is obtained from its prior with the remaining elements 
of $\ugamma_{jk}$, denoted by $\ugamma_{Cjk}$, kept at their current value.   
The proposal pmf is obtained from the Bernoulli prior with $\pi_{\mu j k}$ integrated out
\begin{eqnarray}
p(\ugamma_{Bjk}|\ugamma_{Cjk}) = \frac{p(\ugamma_{jk})}{p(\ugamma_{Cjk})}= 
\frac{\text{Beta}(c_{\mu jk}+N(\ugamma_{jk}),d_{\mu jk}+q_{\mu k}-N(\ugamma_{jk}))}
{\text{Beta}(c_{\mu j k}+N(\ugamma_{Cjk}),d_{\mu jk}+q_{\mu k}-L(\ugamma_{Bjk})-N(\ugamma_{Cjk}))},\nonumber
\end{eqnarray}
where $L(\ugamma_{Bjk})$ denotes the length of $\ugamma_{Bjk}$ i.e. the size of the block. 
For this proposal pmf, the acceptance probability of the Metropolis-Hastings move reduces
to the ratio of the likelihoods in (\ref{marginal}) of the main body of the paper
\begin{eqnarray}
\min\left\{1,(c_{\beta}+1)^{\{N(\gamma^C) - N(\gamma^P)\}/2} \exp\{(S^C-S^P)/2\}\right\}, \nonumber
\end{eqnarray}
where superscripts $P$ and $C$ denote proposed and currents values respectively. 


\item 

Pairs $(\udelta_{jk}, \ualpha_{jk}), j=1,\dots,p, k=1,\dots,Q,$ are updated simultaneously.
Similarly to the updating of $\ugamma_{jk}$, the elements of $\udelta_{jk}$ are updated in random 
order and in blocks of random size. Let $\udelta_{Bjk}$ denote a block. Blocks $\udelta_{Bjk}$ and the whole vector
$\ualpha_{jk}$ are generated simultaneously. As was mentioned by \citet{Chan06}, generating the whole
vector $\ualpha_{jk}$, instead of subvector $\ualpha_{Bjk}$, is necessary
in order to make $\ualpha_{jk}$ consistent with the proposed value of $\udelta_{jk}$. 

Generating the proposed value for $\udelta_{Bjk}$ is done in a similar way as was done for $\ugamma_{Bjk}$. 
Let $\udelta^P_{jk}$ denote the proposed value of $\udelta_{jk}$. Next, we describe how  
the proposed vale for $\ualpha_{\delta^P_{jk}jk}$ is obtained. 
To avoid clutter, proposed values $\ualpha^P_{\delta^P_{jk}jk}$ will be denoted by the simpler $\ualpha_{jk}^P$.  
The development that follows is in the spirit of \citet{Chan06} who built on the work of \citet{Gamerman1997}. 

Let $\hat{\ubeta}_{\gamma}^C = \{c_{\beta}/(1+c_{\beta})\}(\utX_{\gamma}^{\top} \utX_{\gamma})^{-1}\utX_{\gamma}^{\top} \utY$
denote the current value of the posterior mean of $\ubeta_{\gamma}$.
Define the current squared residuals 
\begin{equation}
e_{ij}^C = (y_{ij} - \hat{\beta}_{0j}^C - \ux_{\gamma_j i}^{\top} \hat{\ubeta}_{\gamma_j j}^C)^2. \nonumber
\end{equation}
These have an approximate $\sigma^2_{ij} \chi^2_1$ distribution,
where $\sigma^2_{ij} = \sigma^2_{j} \exp(\uz_{\delta_j i}^{\top} \ualpha_{\delta_j j})$. 
The latter defines a Gamma generalized linear model (GLM) for the squared 
residuals with mean $\sigma^2_{ij}$, which,
utilizing a $\log$-link, can be thought of as Gamma GLM with an 
offset term: $\log(\sigma^2_{ij}) = \log(\sigma^2_{j}) + \uz_{\delta_j i}^{\top} \ualpha_{\delta_j j}$.
Given $\udelta^P_{jk}$, the proposal density for $\ualpha_{\delta^P_{jk}jk}$ 
is derived utilizing the one step iteratively reweighted least squares algorithm.
This proceeds as follows. First define the transformed observations
\begin{eqnarray}
d_{ij}^C(\ualpha^C_j) = \log(\sigma^2_{j}) + \uz_i^{\top} \ualpha_{j}^C + 
\frac{e_{ij}^C-(\sigma^2_{ij})^C}{(\sigma^2_{ij})^C},\nonumber
\end{eqnarray}
where superscript $C$ denotes current values. Further, let $\ud_j^C$ denote the vector of $d_{ij}^C$.

Next we define
\begin{eqnarray}
\uDelta(\udelta^P_{jk}) = (c_{\alpha j}^{-1}\uI + \uZ_{\delta^P_{jk}}^{\top}\uZ_{\delta^P_{jk}})^{-1}
\text{\;and\;}
\ahat(\udelta^P_{jk},\ualpha^C_j) =  \uDelta_{\delta^P_{jk}} \uZ_{\delta^P_{jk}}^{\top} \ud_j^C, \nonumber
\end{eqnarray}
where $\uZ_{\delta_{jk}}$ is a submatrix of $\uZ_{\delta_j}$ that was defined after (\ref{mv1}), and it considers only
the columns that pertain to the $k$th effect.
The proposed value $\ualpha_{jk}^P$ is obtained from a multivariate normal
distribution with mean $\ahat(\udelta^P_{jk},\ualpha^C_j)$ and covariance $h \uDelta(\udelta^P_{jk})$,
denoted as 
$N(\ualpha_{jk}^P;\ahat(\udelta^P_{jk},\ualpha^C_{j}),h_{jk} \uDelta(\udelta^P_{jk}))$,
where $h_{jk}$ is a free parameter that 
we introduce and select adaptively \citep{roberts_examples_2009} in order to achieve an acceptance probability 
of $20\% - 25\%$ \citep{Roberts2001c}.

Let $N(\ualpha_{jk}^C;\ahat(\udelta^C_{jk},\ualpha^P_j),h_{jk}\uDelta(\udelta^C_{jk}))$ denote the proposal density 
for taking a step in the reverse direction, from model $\udelta^P_{jk}$ to $\udelta^C_{jk}$. 
Then the acceptance probability of the pair $(\udelta^P_{jk},\ualpha^P_{\delta^P_{jk}jk})$ is  
\begin{eqnarray}
\min\left\{1,
\frac{|\uSigma(\uR,\ualpha^P,\udelta^P,\usigma^2)|^{-\frac{1}{2}} \exp(-S^P/2)}
{|\uSigma(\uR,\ualpha^C,\udelta^C,\usigma^2)|^{-\frac{1}{2}} \exp(-S^C/2)}
\frac{
	(2\pi c_{\alpha j})^{-\frac{N(\delta^P_{jk})}{2}}\exp\{-\frac{1}{2c_{\alpha j}} (\ualpha^P_{jk})^{\top} \ualpha^P_{jk}\}
}{
	(2\pi c_{\alpha j})^{-\frac{N(\delta^C_{jk})}{2}}\exp\{-\frac{1}{2c_{\alpha j}} (\ualpha^C_{jk})^{\top} \ualpha^C_{jk}\}
}
\frac{
	N(\ualpha_{jk}^C;\ahat_{\delta^C_{jk}},h_{jk} \uDelta_{\delta^C_{jk}})
}{
	N(\ualpha_{jk}^P;\ahat_{\delta^P_{jk}},h_{jk} \uDelta_{\delta^P_{jk}})
}
\right\},\nonumber
\end{eqnarray}
where the determinants, for centred variables, are equal to one, otherwise, the ratio of the determinants may be computed as
$\prod_{i=1}^n\{(\sigma^2_{ij})^{C}/(\sigma^2_{ij})^{P}\}^{1/2}$.


\item 

The full conditional of $\sigma^2_{j}, j=1,\dots,p,$ is given by
\begin{eqnarray}
f(\sigma^2_{j}|\dots) \propto |\uSigma(\uR,\ualpha,\udelta,\usigma^2)|^{-\frac{1}{2}} \exp(-S/2) \xi(\sigma^2_{j}),\nonumber
\end{eqnarray}
where $\xi(\sigma^2_{j})$ denotes either the IG or half-normal prior.
We follow a random walk algorithm obtaining proposed values $(\sigma^2_{j})^{(P)} \sim N((\sigma^2_{j})^{(C)},f_{3j}^2)$, where $f_{3j}^2$ is a tuning parameter that we choose adaptively \citep{roberts_examples_2009} in order
to achieve an acceptance probability of $20\% - 25\%$ \citep{Roberts2001c}. 
Proposed values are accepted with probability $f((\sigma^2_{j})^{(P)}|\dots) /f((\sigma^2_{j})^{(C)}|\dots)$, which reduces to
\begin{eqnarray}
\{(\sigma^2_{j})^{C}/(\sigma^2_{j})^{P}\}^{n/2} \exp\{(S^C-S^P)/2\}) \xi((\sigma^2_{j})^{(P)})/\xi((\sigma^2_{j})^{(C)}).\nonumber
\end{eqnarray}

\item Parameter $c_{\beta}$ is updated from the marginal (\ref{marginal}) and the IG$(a_{\beta},b_{\beta})$ prior 
\begin{eqnarray}
f(c_{\beta}|\dots) \propto (c_{\beta}+1)^{-\frac{N(\gamma)+p}{2}} \exp(-S/2)
(c_{\beta})^{-a_{\beta}-1} \exp(-b_{\beta}/c_{\beta}).\nonumber
\end{eqnarray}
To sample from the above, we utilize a normal approximation. Let
$\ell(c_{\beta}) = \log\{f(c_{\beta}|\dots)\}$. We utilize a normal proposal density 
$N(\hat{c}_{\beta},-g^2/\ell^{''}(\hat c_{\beta}))$ where $\hat{c}_{\beta}$ is the mode
of $\ell(c_{\beta})$, found using a Newton-Raphson algorithm, 
$\ell^{''}(\hat c_{\beta})$ is the second derivative of $\ell(c_{\beta})$ evaluated at the mode, 
and $g^2$ is a tuning variance parameter that we choose adaptively \citep{roberts_examples_2009}
to achieve an acceptance probability of $20\% - 25\%$ \citep{Roberts2001c}.
With superscripts $P$ and $C$ denoting proposed and currents values, the acceptance probability 
is the minimum between one and
\begin{equation}
\frac{f(c_{\beta}^{P}|\dots)}{f(c_{\beta}^{C}|\dots)}
\frac{N(c_{\beta}^{C};\hat{c}_{\beta},-g^2/\ell^{''}(\hat c_{\beta}))}
{N(c_{\beta}^{P};\hat{c}_{\beta},-g^2/\ell^{''}(\hat c_{\beta}))}.\nonumber
\end{equation}

\item Concerning parameter $c_{\alpha j}, j=1,\dots,p$, the full conditional corresponding to the IG$(a_{\alpha j},b_{\alpha j})$ prior is 
another inverse Gamma density IG$(a_{\alpha j} + N(\delta_j)/2,b_{\alpha j}+\ualpha_{\delta_j j}^{\top}\ualpha_{\delta_j j}/2)$.

The full conditional corresponding to the half-normal prior $\sqrt{c_{\alpha j}} \sim N(0,\phi^2_{c_{\alpha j}}) I[\sqrt{c_{\alpha j}}>0]$ is 
\begin{eqnarray}
f(c_{\alpha j}|\dots) \propto c_{\alpha j}^{-N(\delta_j)/2} 
\exp(-\ualpha_{\delta_jj}^{\top}\ualpha_{\delta_jj}/2c_{\alpha j}) \exp(-c_{\alpha j}/2\phi^2_{c_{\alpha j}})I[\sqrt{c_{\alpha j}}>0].\nonumber
\end{eqnarray}
We obtain proposed values $c_{\alpha j}^{(P)} \sim N(c_{\alpha j}^{(C)},f_{2j}^2)$, where $c_{\alpha j}^{(C)}$ denotes the current value. 
Proposed values are accepted with probability $f(c_{\alpha j}^{(P)}|\dots)/f(c_{\alpha j}^{(C)}|\dots)$, 
where $f_{2j}^2$ is a tuning parameter. We select its value adaptively \citep{roberts_examples_2009} so as  
to achieve an acceptance probability of $20\% - 25\%$ \citep{Roberts2001c}.

\item Using likelihood (\ref{full}) and prior (\ref{gprior}), we find the posterior $\ubeta^{\ast}$ to be
\begin{eqnarray}
\ubeta_{\gamma}^{\ast} | \dots \sim N(\frac{c_{\beta}}{1+c_{\beta}}(\utX_{\gamma}^{\top} \utX_{\gamma})^{-1}\utX_{\gamma}^{\top} \utY,
\frac{c_{\beta}}{1+c_{\beta}}(\utX_{\gamma}^{\top} \utX_{\gamma})^{-1}).\nonumber
\end{eqnarray}

\item Update $\uR$ as described in the main body of the paper. 

\item To sample from the full conditional of $\utheta$, write $f(\ur|\utheta,\tau^2) = \nu(\utheta,\tau^2) N(g(\ur);\utheta,\tau^2\uI)$
for the likelihood in (\ref{priorR2}). Further, the prior for $\utheta$ is given in  (\ref{thetaP}), $\utheta \sim N(\mu_R \uone,\sigma^2_R \uI)$. 
Hence, it is easy to show that the posterior is
\begin{eqnarray}\label{ptheta2}
f(\utheta|\dots) = \nu(\utheta,\tau^2)
N\left(\utheta;  \uA (\tau^{-2} g(\ur) + \sigma^{-2}_R \mu_R \uone),\uA \equiv 
(\tau^{-2} + \sigma_{R}^{-2})^{-1}\uI\right).
\end{eqnarray}
At iteration $u+1$ we sample $\utheta^{(u+1)}$ utilizing as proposal the normal distribution
that appears on the right hand side of (\ref{ptheta2}).  
The proposed $\utheta^{(u+1)}$ is accepted with probability
\begin{eqnarray}
\min\left\{1,\frac{\nu(\utheta^{(u+1)},\tau^2)}{\nu(\utheta^{(u)},\tau^2)}\right\},\nonumber
\end{eqnarray}
which, for a small value of $\tau^2$ can reasonably be assumed to be unity
\citep{Liechty, Yu201417,Liechty2}. 

\item Update $\mu_R$ from $\mu_R \sim N((d/\sigma^2_R+1/\varphi^2_r)^{-1} (d/\sigma^2_R)\bar{\theta},(d/\sigma^2_R+1/\varphi^2_r)^{-1})$,
where $\bar{\theta}$ is the mean of the elements of vector $\utheta$. 

\item We update $\sigma^2_R$ utilizing the following full conditional 
\begin{eqnarray}
f(\sigma^2_R|\dots) \propto (\sigma_R^2)^{-\frac{d}{2}} \exp\{-\sum_{i=1}^d(\theta_i-\mu_R)^2 /(2\sigma_R^2)\} \exp\{-\sigma_R^2/(2\phi^2_{R})\} I[\sigma_R>0].\nonumber
\end{eqnarray}
Proposed values are obtained from $(\sigma^2_{R})^{(p)} \sim N((\sigma^2_{R})^{(c)},f_1^2)$ where $(\sigma^2_{R})^{(c)}$ denotes the current value. 
Proposed values are accepted with probability $f((\sigma^2_{R})^{(p)}|\dots)/f((\sigma^2_{R})^{(c)}|\dots)$. 
We treat $f_1^2$ as a tuning parameter and we select its value adaptively \citep{roberts_examples_2009} in 
order to achieve an acceptance probability 
of $20\% - 25\%$ \citep{Roberts2001c}.

\end{enumerate}

\subsection{MCMC algorithm for the grouped correlations model}

With the introduction of the shadow prior, model (\ref{priorR3}) becomes the same as in (\ref{priorR2}). The difference 
is in the distribution of $\theta_{kl},$ which are now independently distributed with conditional distribution 
$\theta_{kl} | \lambda_{kl} = h \sim N(\mu_{R,h}, \sigma^2_R)$. Here we point out the additional MCMC steps needed for the 
`grouped correlations' models:
\begin{enumerate}
	\item Let $\utheta_h$ denote the vector of $\theta_{kl}$ that have been assigned to cluster $h$, $h=1,\dots,H$. 
	The posterior of $\utheta_h$ is
	\begin{eqnarray}
	f(\utheta_h|\dots) \propto
	N\left(\utheta_h;  \uA (\tau^{-2} g(\ur) + \sigma^{-2}_R \mu_{R,h} \uone),\uA \equiv 
	(\tau^{-2} + \sigma_{R}^{-2})^{-1}\uI\right).\nonumber
	\end{eqnarray}
	
	\item Update $\mu_{R,h}$ from $\mu_{R,h} \sim N((d_h/\sigma^2_R+1/\varphi^2_r)^{-1} (d_h/\sigma^2_R)\bar{\theta}_h,(d_h/\sigma^2_R+1/\varphi^2_r)^{-1})$,
	where $d_h$ is the number of $\theta_{kl}$ assigned to the $h$th cluster and $\bar{\theta}_h$ is their mean. 
	
	\item Update $v_h \sim\text{Beta}(d_h+1,d-\sum_{l=1}^hd_h+\alpha^*), h=1,\dots,H-1,$
	where $d_h$ is the number of correlations allocated in the $h$th cluster. 
	Given $v_h,$ update the stick-breaking weights $w_h, h=1,\dots,H$. 
	
	\item Posterior cluster assignment probabilities are computed using  
	\begin{eqnarray}
	P(\lambda_{kl}=h|\dots) \propto w_{h} N(\theta_{kl};\mu_{R,h},\sigma^2_R).\nonumber 
	\end{eqnarray} 
	
	\item We update concentration parameter $\alpha^*$ using the method described by \citet{EscobarWest}.
	With the $\alpha^{*} \sim \text{Gamma}(a_{\alpha*},b_{\alpha*})$ prior,
	the posterior can be expressed as a mixture of two Gamma distributions: 
	\begin{equation}\label{alphadist}
	\alpha^*|\eta,k \sim \pi_{\eta} \text{Gamma}(a_{\alpha*}+k,b_{\alpha*}-\log(\eta)) + (1-\pi_{\eta}) 
	\text{Gamma}(a_{\alpha*}+k-1,b_{\alpha*}-\log(\eta)),
	\end{equation}
	where $k$ is the number of non-empty clusters, 
	$\pi_{\eta} = (a_{\alpha*}+k-1)/\{a_{\alpha*}+k-1+n(b_{\alpha*}-\log(\eta))\}$ and 
	\begin{equation}\label{etadist}
	\eta|\alpha^*,k \sim \text{Beta}(\alpha^*+1,d). 
	\end{equation}
	Hence the algorithm proceeds as follows: with $\alpha^*$ and $k$ fixed at their current values, we sample $\eta$ from
	(\ref{etadist}). Then, based on the same $k$ and the newly sampled value of $\eta$, we sample a new $\alpha$ value
	from (\ref{alphadist}). 
	
\end{enumerate}

\subsection{MCMC algorithm for the grouped variables model}

Here we point out the only difference between the MCMC algorithms for grouped correlations and 
grouped variables models:
\begin{enumerate}
	\item Let $w_{h}$ be the prior probability that a variable is assigned to cluster $h$.
	Then cluster assignment probabilities are computed as follows
	\begin{eqnarray}
	P(\lambda_{k}=h|\dots) \propto 
	w_{h} \prod_{l \neq k} N(\theta_{kl};\mu_{R,h,\lambda_{l}},\sigma^2_R).\nonumber 
	\end{eqnarray} 
\end{enumerate}

\bibliographystyle{apalike}
\bibliography{all.bib}